\documentclass[%
 aip, jcp,
 amsmath,amssymb,
 reprint,%
]{revtex4-2}

\usepackage{graphicx}
\usepackage{dcolumn}

\usepackage{bm}

\usepackage[utf8]{inputenc}
\usepackage[T1]{fontenc}
\usepackage{pifont}
\usepackage{newtxmath, newtxtext}

\DeclareMathAlphabet{\mathcal}{OMS}{cmsy}{m}{n}
\DeclareFontFamily{OT1}{pzc}{}
\DeclareFontShape{OT1}{pzc}{m}{it}{<-> s * [1.100] pzcmi7t}{}
\DeclareMathAlphabet{\mathpzc}{OT1}{pzc}{m}{it}
\usepackage{booktabs}
\usepackage{mathtools}
\usepackage{relsize}
\usepackage{cleveref}
\usepackage{titlesec}
\usepackage{physics}
\usepackage{siunitx}
\usepackage{chemfig}
\usepackage{multirow}
\usepackage{makecell}
\allowdisplaybreaks
\begin{document}

\preprint{AIP/123-QED}

\title{Swinging between shine and shadow: Theoretical advances on thermally-activated vibropolaritonic chemistry (a perspective).}

\author{J. A. Campos-Gonzalez-Angulo}
\affiliation{Department of Chemistry and Biochemistry. University of California San Diego. La Jolla, California 92093, USA}
 \affiliation{Chemical Physics Theory Group. Department of Chemistry, University of Toronto, Toronto, Ontario, M5S 3H6, Canada}%
\author{Y. R. Poh}%
\affiliation{Department of Chemistry and Biochemistry. University of California San Diego. La Jolla, California 92093, USA}%
\author{M. Du}%
\affiliation{Department of Chemistry and Biochemistry. University of California San Diego. La Jolla, California 92093, USA}%
\author{J. Yuen-Zhou}
 \email{joelyuen@ucsd.edu}
 \homepage{http://yuenzhougroup.ucsd.edu}
\affiliation{Department of Chemistry and Biochemistry. University of California San Diego. La Jolla, California 92093, USA}%

%
\begin{abstract}
Polariton chemistry has emerged as an appealing branch of synthetic chemistry that promises mode selectivity and a cleaner approach to kinetic control. Of particular interest are the numerous experiments in which reactivity has been modified by virtue of performing the reaction inside infrared optical microcavities in the absence of optical pumping; this effort is known as "vibropolaritonic chemistry." The optimal conditions for these observations are (1) resonance between cavity and reactive modes at normal incidence ($k=0$), and (2) monotonic increase of the effect with the concentration of emitters in the sample. Importantly, vibropolaritonic chemistry has only been experimentally demonstrated in the so-called "collective" strong coupling regime, where there is a macroscopic number of molecules (rather than a single molecule) coupled to each photon mode of the microcavity. Strikingly, efforts to understand this phenomenon from a conceptual standpoint have encountered several roadblocks and no single, unifying theory has surfaced thus far. This perspective documents the most relevant approaches taken by theorists, laying out the contributions and unresolved challenges from each work. We expect this manuscript to not only serve as a primer for experimentalists and theorists alike, but also inform future endeavors in the quest for the ultimate formalism of vibropolaritonic chemical kinetics.
\end{abstract}

\maketitle

\section{Introduction\label{sec:intro}}

\begin{figure}
\centering
\includegraphics[width=\linewidth]{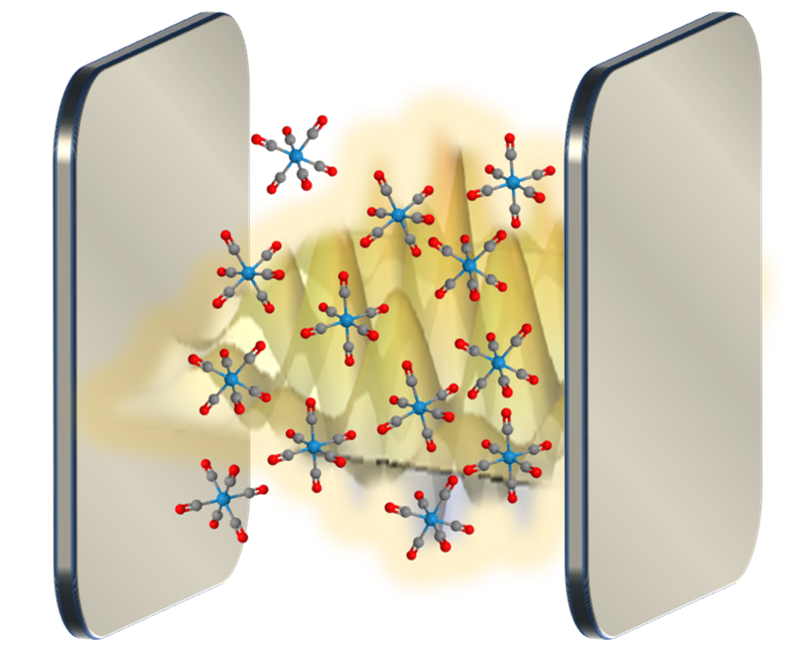}
\caption[Fabry-Pérot resonator hosting organic molecules.]{Fabry-Pérot resonator hosting  a molecular solution.\label{fig:cav1}}
\end{figure}

For decades, chemists have used electromagnetic (EM) radiation to investigate the microscopic structure and properties of substances. In a typical spectroscopy experiment, an EM field interacts with a piece of matter; comparing the field before and after the interaction informs the energetic makeup of the material. In these studies, the light-matter interaction is assumed to be weak enough that the energetic landscape of the substance remains unperturbed by the EM probe. This assumption is valid as long as the energy exchange between the material and the EM field is slower than the dissipation of the radiation. Therefore, confining the EM field in a region of space where the material repeatedly reabsorbs it strengthens the light-matter interaction. This scenario takes place in devices known as optical cavities or resonators (\cref{fig:cav1}), which can be tailored to tune the captured EM mode to a given transition in the energetic spectrum of the substance. Under these conditions, a degree of freedom (e.g., electronic, vibrational, rotational) of the material couples through its transition dipole moment to the EM field, giving rise to excitations with a hybrid light-matter character known as polaritons. The signature of polariton formation is the observation of the Rabi splitting in linear optical spectroscopy (reflection, transmission or absorption), i.e., the resolution of the resonant spectral signals into well-defined peaks centered away from the resonant frequency. The peak-to-peak distance (Rabi frequency) between the polaritonic resonances depends on the light-matter coupling magnitude.

The peculiarities of polaritonic systems, such as a modified energy spectrum and the imprint of photonic character to the material degrees of freedom (DOF), have been exploited for purposes as diverse as Bose-Einstein condensation,\cite{Kasprzak2006} quantum computing,\cite{Blais2004} room temperature lasing,\cite{KenaCohen2010,Sannikov2019} nonlinear optical response,\cite{Barachati2015,Barachati2018} reversible optical switching,\cite{Schwartz2011} enhanced charge conductivity,\cite{Orgiu2015,Kang2021} and long-range excitation energy transfer.\cite{Zhong2017,Georgiou2018} Theoretical considerations on the latter phenomenon \cite{MartinezMartinez2018,Du2018} have led to the proposal of intriguing ideas such as remote catalysis.\cite{Du2019}

Recently, there have been remarkable efforts to control chemical reactivity with cavity resonances.\cite{Ebbesen2016,Ribeiro2018,Feist2018,Flick2018,KenaCohen2019,Herrera2020,Hirai2020} In Ref. \onlinecite{Hutchison2012}, Hutchison and co-workers demonstrated for the first time an observable effect of strong light-matter coupling on a chemical reaction. Specifically, the authors observed a Rabi splitting-dependent slowdown in the photoisomerization of spiropyran to merocyanine when the latter was coupled to the cavity. Another example of cavity-modified excited-state chemistry can be found in Ref. \onlinecite{Munkhbat2018}, where Munkhbat and collaborators observed suppression of photo-oxidation for a dye aggregate by tuning the cavity to the exciton frequency of the sample. These and other excited-state chemical processes \cite{Stranius2018,Eizner2019} have motivated an intense theoretical interest in the field of polaritonic chemistry \cite{Galego2015,Herrera2016,MartinezMartinez2018,Flick2017,Flick2017a,Semenov2019} and have been, to a certain extent, understood based on how light-matter coupling modifies the relaxation channels available during photoinitiated processes.\cite{Fregoni2022}

In the pursuit of mode-selective selective reactions, it is reasonable to also explore scenarios where the EM field engages with excitations hosted in molecular bond vibrations. Conveniently, strong coupling in the infrared, hereafter known as vibrational strong coupling (VSC), has been observed for various substances, such as polymers,\cite{Long2015,Shalabney2015} proteins, \cite{Vergauwe2016} organometallic complexes,\cite{Simpkins2015,Casey2016} and, remarkably, organic solutes,\cite{George2015} among others.\cite{Mason2012} VSC has been extensively investigated from both theoretical and experimental fronts. Among the developments along this line of research, there are conjectured changes in Raman scattering cross-sections,\cite{Shalabney2015,Pino2015a,Strashko2016} mode hybridization,\cite{Muallem2016,Crum2018} two-dimensional spectroscopy, \cite{Saurabh201811,Xiang2018} and non-linear response.\cite{Ribeiro2018,Xiang2019,Ribeiro2020}

One of the most striking developments regarding molecules and cavities in the infrared is \emph{vibropolaritonic chemistry}, where it has been observed that performing a chemical reaction inside a resonator can modify the rate of the process even in the absence of external optical pumping.\cite{KenaCohen2019,Hirai2020} This phenomenon is the central topic of the present perspective.

This manuscript joins the wave of recently published perspectives on the topic,\cite{Nagarajan2021,Simpkins2021,Wang2021,YuenZhou2021,Dunkelberger2022,Li2022,Sidler2022} which is a testimony of the currently vibrant state of the field. Despite the surge in experimental evidence,\cite{Nagarajan2021,Ahn2022} conceptualizing this phenomenon in terms of well-established theoretical frameworks, which have been largely successful at addressing the optical response of polaritonic systems, remains an ongoing effort.\cite{Wang2021,Fregoni2022} Specifically, light-matter interaction, which is described by cavity quantum electrodynamics (CQED), and chemical kinetics are well understood when studied independently. Yet, every attempt to merge the toolkits that these two conventional fields provide has resulted in models that either contradict experimental observations (predict no change in the rate), or support it with caveats that are incompatible with experimental settings. Theorists\cite{Climent2020,Thomas2020b,Climent2021} and experimentalists\cite{Imperatore2021,Wiesehan2021} alike have becoming increasingly concerned with this mystery.

The present perspective serves as a detailed summary of the attempts made at explaining how VSC modifies chemical groundstate kinetics. It complements the work in Ref. \onlinecite{Wang2021} not only by updating the state of affairs, but also by presenting more of the quantitative aspects of each approach. By providing a recount of theoretical works, we hope to give a balanced portrayal of the contributions and new considerations that each publication brought into the picture, as well as the specific reasons why every approach remains an incomplete bridge between theory and experiment.

This perspective is organized as follows: First, in Sec. \ref{sec:experim}, we introduce and roughly summarize the experimental results that prompted the exploration of vibropolaritonic chemistry. Next, in Sec. \ref{sec:theos}, we revisit the conventional description of VSC and formulations of chemical rate theory relevant to chemical dynamics. Continuing in Sec. \ref{sec:approaches}, we recount and summarize literature that attempts to explain rate modifications theoretically. Finally, we present a discussion and outlook in Sec. \ref{sec:conc}.

\section{Cavity-induced modifications of ground-state chemical kinetics: Experimental observations of vibropolaritonic chemistry\label{sec:experim}}

Since 2016, the Nanostructures Laboratory at the Université de Strasbourg, directed by Prof. Thomas Ebbesen, has led the charge in producing a train of experimental evidence suggesting that the kinetics of a chemical reaction is modified when it takes place inside a Fabry-Pérot (FP) microcavity, under conditions consistent with VSC.

For instance, in Ref. \onlinecite{Thomas2016}, Thomas and co-workers observed that the deprotection of 1-phenyl-2-trimethyl-silylacetylene experiences slowdown when performed in a FP cavity with inter-mirror separation tuned to produce the maximum Rabi splitting over the peak at $\SI{860}{\per\centi\metre}$ at normal incidence ($k=0$); this peak is arguably assigned to the stretching mode of the \chemfig[atom sep=2em]{Si-C} bond that breaks during the reaction.\cite{Thomas2020} They found that varying the length of the cavity gap reduces the deviation between the rates measured inside and outside of the cavity, following a trend reminiscent of the absorption peak, i.e., the effect was maximum at resonance and decreased with the detuning following the absorption line shape. Furthermore, they found a positive correlation between the magnitude of the rate deceleration and the reactant concentration; this observation is interpreted as a consequence of the rate being dependent on the Rabi splitting. They also extracted kinetic parameters and found that the cavity created an effective increase in the activation enthalpy and a transition from negative to positive effective activation entropy. The latter suggests a phenomenological change in the reaction mechanism, going from a bimolecular to a unimolecular rate determining step.\cite{Thomas2020} It is important to note that, while variations in cavity length reduce VSC at $k=0$, they enhance VSC at higher values of $k$ (other photon modes at oblique incidence), so it is perplexing that vibropolaritonic chemistry occurs only for VSC at $k=0$ but not at other $k>0$ (the same phenomenon is observed throughout the other vibropolaritonic chemistry experiments listed below).

Ebbesen's research group built upon these findings and published a study with a similar reaction but in which the reactant, \emph{tert}-butyldimethyl\{[4-(trimethylsilyl)but-3-yn-1-yl]oxy\}silane, has two labile sites: \chemfig[atom sep=2em]{Si-C} ($\SI{842}{\per\centi\metre}$) and \chemfig[atom sep=2em]{Si-O} ($\SI{1110}{\per\centi\metre}$).\cite{Thomas2019} This time, they tuned the cavity to the bending mode of \chemfig[atom sep=2em]{Si-CH_3} ($\SI{1250}{\per\centi\metre}$) and the stretching mode of \chemfig[atom sep=2em]{C-O} ($\SI{1045}{\per\centi\metre}$); these bonds do not break during the reactions. Unsurprisingly, they found that coupling to the \chemfig[atom sep=2em]{C-O} bond does not affect the rate. In contrast, reactivity deceleration is observed whenever any of the remaining three modes engages with the cavity, with the identity of the coupled mode impacting only slightly the magnitude of the effect. Furthermore, the scission of \chemfig[atom sep=2em]{Si-C} is more affected than that of \chemfig[atom sep=2em]{Si-O}, independent of the coupled mode. This imbalance in the reaction slowdown enables an inversion of the branching ratio (between the two different products) upon VSC. Clearly the authors did not achieve their desired selectivity of the coupled mode being affected above other modes. Despite that, the findings of this study have inspired a development of strategies towards mode-selective chemistry.

Resonant decelerating effects that increase with the Rabi splitting have also been observed in prins cyclization \cite{Hirai2020} and the proteolytic activity of pepsin.\cite{Vergauwe2019} In the former case, resonance was achieved between the \chemfig[atom sep=2em]{C=O} stretch and the cavity, whereas, in the latter case, the solvent molecules coupled resonantly to the cavity. A variety of reactions were also performed under the so-called ultrastrong coupling regime,\cite{Hiura2018} i.e., with Rabi splitting larger than 10 \% of the resonant cavity frequency. Moderate rate increases were reported for nucleophilic additions of isocyanates and cyanate ion, as well as ketene cycloadditions involving \chemfig[atom sep=2em]{Ph_{3}P=C=C=O}; this is in contrast with the larger rate accelerations of up to four orders of magnitude observed in the hydrolysis of cyanate ion and ammonia borane. Coupling to the solvent molecules has also been explored experimentally. For instance, moderate enhancements were observed with the solvolysis of the ester group in \emph{p}-nitrophenyl acetate when the \chemfig[atom sep=2em]{C=O} stretch of the solvent, ethyl acetate, was strongly coupled.\cite{Lather2019} Interestingly, by substituting the carbonyl \chemfig{^{12}C} of ethyl acetate with \chemfig{^{13}C}, the authors observed a secondary kinetic isotope effect that was stronger inside the cavity than outside, suggesting that VSC influences the reaction coordinate. Moreover, the transesterification of \emph{p}-nitrophenyl bezoate derivatives experienced catalysis in cavities resonant with the stretching band corresponding to \chemfig[atom sep=2em]{C=O} in both the reactant and the solvent, isopropyl acetate ($\SI{1739}{\per\centi\metre}$).\cite{Lather2022} In these reactions, the free-energy relationship given by the Hammett plot (of relative reaction rate vs. equilibrium constant) was no longer linear when the system was under VSC.

A prominent step forward towards the utilization of VSC in mode selective chemistry can be found in Ref. \onlinecite{Pang2020}. There, Pang and collaborators coupled several vibrational modes of mesitylene and other charge transfer donors to resonant cavity modes and measured the equilibrium constant of the complexation reaction with iodine. They found that coupling to modes with the $A'$ irreducible representation of the $C_{3h}$ symmetry group resulted in deceleration, while coupling to $E'$ modes produced acceleration.

 All these results, summarized from more experimentally oriented perspectives in Refs. \onlinecite{Hirai2020,Nagarajan2021}, indicate that the EM mode is effectively modifying the energetic landscape of the ground electronic state, consistent with some of the notions of polariton formation. However, despite several laboratories generating these results, other experimental groups have highlighted difficulties in measuring those kinetic effects using prescriptions from the Ebbesen school, from which most successful vibropolaritonic chemistry experiments originated.\cite{Imperatore2021,Simpkins2021,Wiesehan2021} This might change though; recently, a preprint by Ahn at Bilkent University, Herrera at the Universidad de Santiago de Chile and Simpkins at the U.S. Naval Research Laboratory reported VSC-mediated rate suppressions for the nucleophilic addition of cyclohexanol to phenyl isocyanate using tetrahydrofuran as the solvent.\cite{Ahn2022} By separately tuning the cavity to the \chemfig[atom sep=2em]{N=C=O}, \chemfig[atom sep=2em]{C-H} and \chemfig[atom sep=2em]{C=O} stretching modes, the researchers were able to investigate, in the same system, resonant couplings to the reactant, solvent and product modes respectively, with the first case giving the largest rate deceleration of $\sim$ 80 \%. However, it is worth noting that, unlike most reactions earlier studied, this reaction has a lower activation barrier of less than twice the reactive vibrational frequency (in fact, in an independent experiment performed outside of a cavity, it was shown to be catalyzed by infrared laser excitation of the reactive mode\cite{Stensitzki2018}).

Needless to say, attaining rational vibropolaritonic chemistry would entail a revolution in all of chemistry as it would provide an ultimate form of heterogeneous catalysis. Because of this reason, the development of a conceptual framework under which the architectures of cavities and the optimal reaction conditions can be designed has become a prominently appealing endeavor. This framework should ideally model the following \textit{conditions} under which vibropolaritonic chemistry experiments were conducted:\cite{YuenZhou2021}

\renewcommand\theenumi{$\mathit{C\arabic{enumi}}$}
\begin{enumerate}
	\item The presence of $N \approx 10^6 - 10^{12}$ molecules \emph{collectively} coupled to the cavity;
	\item The lack of optical pumping, that is, the reaction is driven by \emph{thermal fluctuations},
\end{enumerate}

and display the following experimental \textit{observations}:

\renewcommand\theenumi{$\mathit{O\arabic{enumi}}$}
\begin{enumerate}
	\item The possibility of both \emph{enhancement} and \emph{suppression} of reaction rates by the cavity;
	\item The fact that optimum rate modifications occur when the cavity is in \emph{resonance} with the reactant, spectator and/or solvent vibrational modes, and,
	\item specifically, this resonance occurs with the $k=0$ cavity mode (normal incidence).
\end{enumerate}

\section{Theoretical background\label{sec:theos}}

This section presents the theoretical frameworks under which chemical rates are typically understood and formulates the most compatible description of VSC for its incorporation in the rate framework.\cite{CamposGonzalezAngulo2019,Yang2021}
\begin{figure}
\includegraphics[width=\linewidth]{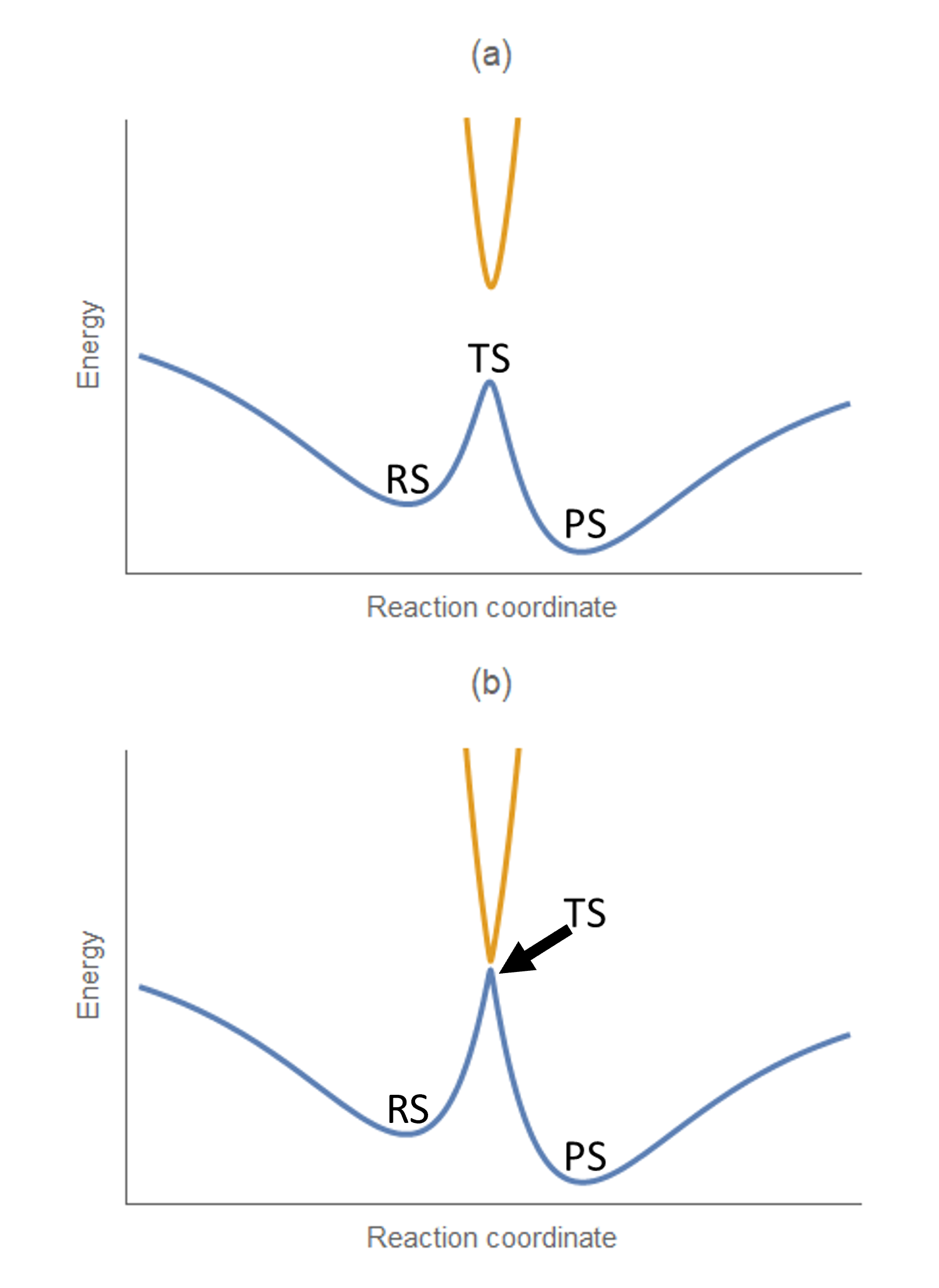}
\caption{Ground and first-excited electronic PESs along the reaction coordinate of (a) an adiabatic reaction and (b) a non-adiabatic reaction.\label{fig:adiapes}}
\end{figure}

\subsection{Transition-state theory\label{sec:TST}}

Most of the reactions for which vibropolaritonic chemical behavior has been reported correspond to the variety of adiabatic transitions. In these processes, the nuclear rearrangements along the reaction coordinate occur slowly enough for the system to always be in the adiabatic electronic ground state, i.e., the Born-Oppenheimer (BO) approximation holds. Then, the reaction may be modelled by nuclei moving along the adiabatic ground potential energy surface (PES), which smoothly connects the reactant state (RS) potential well to the product state (PS) potential well, through a saddle point known as the transition state (TS) (\cref{fig:adiapes}a). Mathematically, within the BO approximation, the Hamiltonian describing the nuclear motion of a molecule with $M$ nuclei is of the form
\begin{equation}
\hat{H}_\textrm{mol}=\sum_{n=1}^M\frac{\hat{P}_i^2}{2}+V_\textrm{mol}(\{\hat{R}_n\}),
\end{equation}
where $\{\hat{P}_n,\hat{R}_n\}$ are the mass-scaled momentum and position operators of the $n$th nucleus, and $V_\textrm{mol}(R)$ is the potential energy function, which, to a good approximation, can be split into translational, rotational, and vibrational contributions, i.e., $V_\textrm{mol}=V_\textrm{tra}+V_\textrm{rot}+V_\textrm{vib}$. For the purpose of reaction dynamics, we shall focus on $V_\textrm{vib}$. When written in terms of the eigenvalues of $\{\hat{R}_n\}$, the term $V_\textrm{vib}(\{R_n\})$ is known as the PES, which, for an adiabatic reaction, has two local minima identified as RS and PS that are connected through the TS saddle point. Around the RS and TS (known generally as critical points), it is possible to define coordinate transformations $\{R_n\}\to\{q_{r\zeta}\}$ where $q_{r\zeta}$ denotes the elongation (distance from the critical point) along the $\zeta$th normal mode, and $r$ labels the stationary point, i.e.,
\[q_{r\zeta}\to
\begin{cases}
q_{\textrm{eq}\zeta}&\text{if original }\{R_n\}\text{ were around RS},\\
q_{\ddagger\zeta}&\text{if original }\{R_n\}\text{ were around TS}.
\end{cases}\]
Then, the PES in the neighborhood of the critical point can be written approximately as a sum over the harmonic oscillator potentials of each normal mode coordinate, i.e.,
\begin{equation}
V_\textrm{vib}(\{q_{r\zeta}\})= V_r+\sum_{\zeta=1}^\Gamma\frac{\omega_{r\zeta}^{2}}{2}q_{r\zeta}^{2}+\order{q_{r\zeta}^{3}},
\end{equation}
where $V_r=V_\textrm{vib}(\{q_{r\zeta}=0\})$ is the value of the potential at the critical point, $\omega_{r\zeta}^{2}=\qty(\partial^2{V_\textrm{vib}}/\partial{q_{r\zeta}}^2)_{q_{r\zeta}=0}$ corresponds to the frequency of the $\zeta$th oscillator defined by the convexity (concavity) at the critical point, and $\Gamma$ is the number of vibrational modes that the molecule has ($3M-5$ for a linear molecule and $3M-6$ for a nonlinear molecule). Note that, since the TS is a saddle point, one of the normal modes in its neighborhood is unstable, that is, motion along this coordinate corresponds to a maximum. This is represented by an imaginary frequency along this mode, i.e., $\omega_{\ddag\zeta}^{2}<0$ for the unstable (reactive) mode $\zeta=\Gamma$ and $\omega_{\ddag\zeta}^{2}>0$ for all other modes.
 
 When the above description suits the energetic landscape, the reaction rate may be computed by transition-state theory (TST) as\cite{Wigner1938,Haenggi1990,Pollak2005,Arnaut2006,Henriksen2018}
 \begin{equation}\label{eq:rateconst}
k_\textrm{TST}=\frac{k_\textrm{B}T}{2\pi\hbar}\frac{Q_\ddag^\textrm{(tra,rot)}Q_\ddag^\textrm{(vib)}}{Q_\textrm{eq}^\textrm{(tra,rot)}Q_\textrm{eq}^\textrm{(vib)}}\textrm{e}^{-E_a/k_\textrm{B}T},
\end{equation}
with $\hbar$ and $k_\textrm{B}$ being the Planck's reduced and Boltzmann's constants, respectively, and $T$ being the temperature. Also, $E_a=V_\ddag+\sum_{\zeta=1}^{\Gamma-1}\hbar\omega_{\ddag\zeta}/2-V_\textrm{eq}-\sum_{\zeta=1}^{\Gamma}\hbar\omega_{\textrm{eq}\zeta}/2$ denotes the activation energy, where $V_{\ddagger}-V_{\textrm{eq}}$ is the TS energy relative to the RS energy and the remaining terms encompass the zero-point energy (ZPE) effects.

 In \cref{eq:rateconst}, the terms $Q_r^\textrm{(tra,rot)}$ encapsulate the contributions of rotational and translational DOF to the partition function, while $Q_r^\textrm{(vib)}$ refers to the vibrational DOF. Explicitly, the contribution to the rate from the vibrational DOF is given by
\begin{equation}
\frac{Q_\ddag^\textrm{(vib)}}{Q_\textrm{eq}^\textrm{(vib)}}=Q_\textrm{vib}^{-1}(\omega_{\textrm{eq}\Gamma})\prod_{\zeta=1}^{\Gamma-1}\frac{Q_\textrm{vib}^{-1}(\omega_{\textrm{eq}\zeta})}{Q_\textrm{vib}^{-1}(\omega_{\ddag\zeta})},
\end{equation}
where $Q_\textrm{vib}^{-1}(\omega)=2\sinh(\hbar \omega/2k_\textrm{B} T)$. Note that the indexing inside the product is arbitrary and $\omega_{\ddag\zeta}$ is generally unrelated to $\omega_{\textrm{eq}\zeta}$. In the high-temperature limit, $Q_\textrm{vib}^{-1}(\omega)\approx\hbar\omega/k_BT$, and\cite{Nitzan2006}
\begin{equation}\label{eq:partfn}
\frac{Q_\ddag^\textrm{(vib)}}{Q_\textrm{eq}^\textrm{(vib)}}\approx\frac{\hbar\omega_{\textrm{eq}\Gamma}}{k_BT}\prod_{\zeta=1}^{\Gamma-1}\frac{\omega_{\textrm{eq}\zeta}}{\omega_{\ddag\zeta}}.
\end{equation}
If a single coordinate is enough to describe the reaction ($\Gamma=1$), i.e., all the non-reactive normal modes are identical in the RS and TS, we have $Q_\ddag^\textrm{(vib)}/Q_\textrm{eq}^\textrm{(vib)}=Q_\textrm{vib}^{-1}(\omega_\textrm{eq})$.

When phenomena absent in the above standard TST play a determining role on the rate constant, their effects are customarily introduced in the form of a transmission coefficient $\kappa$, such that the observed rate constant is $k_\textrm{obs}=\kappa k_\textrm{TST}$.

\subsection{Charge transfer rate}

\begin{figure}
\includegraphics[width=\linewidth]{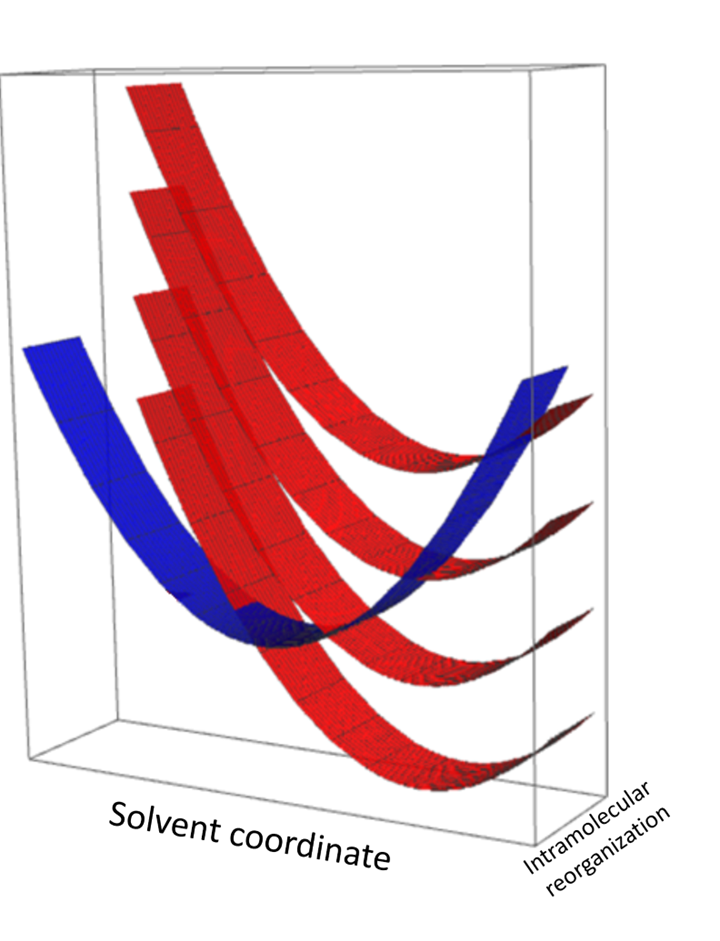}
\caption{Representation of PESs in the MLJ model of electron transfer. The blue curve represents the reactant global ground state while each of the red curves corresponds to high-frequency vibrational excitations of the product electronic ground state. \label{fig:mljpes}}
\end{figure}

In contrast with adiabatic transitions, non-adiabatic processes occur so fast that the system is unable to adjust its configuration to the adiabatic electronic ground state, that is, the nuclear rearrangement leading to the PS promotes the system to an excited adiabatic electronic state. From another perspective, the system stays mostly in the diabatic PES of the RS and rarely crosses over to the PS (\cref{fig:adiapes}b). In this scenario, it is convenient to approximate the diabatic PES of the PS as a harmonic oscillator displaced from that of the RS, and the reaction corresponds to a perturbative electronic interaction between the two PESs. When this description is adequate, the rate constant of the process is accurately calculated with Fermi\rq{}s golden rule. Furthermore, if the system can be separated into low-frequency modes related to the intermolecular DOF, and high-frequency modes describing the intramolecular vibrations, then the former admit a semiclassical treatment, while the latter require a quantum portrayal.\cite{Islampour1993,Heller2020} More concretely, the rate constant is prescribed by the Marcus-Levich-Jortner (MLJ)\cite{Marcus1964,Levich1966,Jortner1975} theory for electron transfer as\cite{May2011,Chaudhuri2017}
\begin{subequations}
\begin{equation}\label{eq:MLJrate}
k_\textrm{MLJ}=\frac{1}{\hbar}\sqrt{\frac{\pi}{k_BT}}\sum_{(\varphi\rq{},\bm{\zeta}\rq{})\neq(\varphi,\bm{\zeta})}\exp\qty(-\frac{E_{\varphi,\bm{\zeta}}}{k_B T})k^{(\varphi\rq{},\bm{\zeta}\rq{})}_{(\varphi,\bm{\zeta})},
\end{equation}
with
\begin{equation}
k^{(\varphi\rq{},\bm{\zeta}\rq{})}_{(\varphi,\bm{\zeta})}=\frac{\abs{J_{\varphi\varphi\rq{}}}^2}{\sqrt{\lambda_S^{(\varphi\varphi\rq{})}}}\abs{F^{(\varphi\rq{},\bm{\zeta}\rq{})}_{(\varphi,\bm{\zeta})}}^2\exp(-\frac{{E_a}^{(\varphi\rq{},\bm{\zeta}\rq{})}_{(\varphi,\bm{\zeta})}}{k_BT}),
\end{equation}
\end{subequations}
where $\varphi$ labels an electronic state while $\bm{\zeta}$ indicates the set of excitations in the vibrational normal modes belonging to $\varphi$. Primed indices tag the products, while unprimed the reactants. The symbol $\lambda_S^{(\varphi\varphi\rq{})}$ denotes the reorganization energy for the low-frequency transition, and $J_{\varphi\varphi\rq{}}$ corresponds to the diabatic coupling between electronic states $\varphi\rq{}$ and $\varphi$. The quantity $F^{(\varphi\rq{},\bm{\zeta}\rq{})}_{(\varphi,\bm{\zeta})}=\braket{\bm{\zeta}(\varphi)}{\bm{\zeta}\rq{}(\varphi\rq{})}$ is the so-called Franck-Condon factor that accounts for the overlap between vibrational eigenfunctions with quantum numbers $\bm{\zeta}$ and $\bm{\zeta}\rq{}$. ${E_a}^{(\varphi\rq{},\bm{\zeta}\rq{})}_{(\varphi,\bm{\zeta})}=(E_{\varphi\rq{},\bm{\zeta}\rq{}}-E_{\varphi,\bm{\zeta}}+\lambda_S^{(\varphi\varphi\rq{})})^2/4\lambda_S^{(\varphi\varphi\rq{})}$ is the activation energy corresponding to the intersection of the parabolic PES defined by the indicated quantum numbers. Within this model, every transition between each pair of vibrational configurations represents a reactive channel (\cref{fig:mljpes}).

For the sake of consistency, this manuscript reports results from theoretical approaches in the form of a transmission coefficient $\kappa$, which is to be interpreted as  $k_\textrm{VSC}/k_\textrm{bare}$, i.e., the reaction rate under VSC relative to the bare rate. The latter term $k_\textrm{bare}$ can be either the TST rate $k_\textrm{TST}$ or the MLJ rate $k_\textrm{MLJ}$, depending on the nature of the rate theory.

\subsection{Description of VSC}\label{sec:theopol}

To understand how the presence of a cavity mode may impact the rate constant, we introduce the formalism under which light-matter coupling is typically explored. To begin with, experiments of VSC are usually performed inside Fabry-Pérot resonators (\cref{fig:cav1}), which allows the frequency of the confined photon mode at $k=0$ to be tuned by varying the distance between two mirrors. Therefore, the confinement volume is reasonably approximated by
\begin{equation}\label{eq:modevol}
\mathcal{V}=\lambda_0\mathcal{S},
\end{equation}
where $\lambda_0$ is the wavelength of the confined EM field and $\mathcal{S}$ is the, presumably fixed, effective cross section area of the confinement volume.  The experimentally adjusted mode volume can thus be written in terms of the cavity mode frequency, $\omega_0$, as $\mathcal{V}=2\pi\sqrt{\varepsilon/\varepsilon_0}c_0\mathcal{S}/\omega_0$, where $\varepsilon$ and $\varepsilon_0$ are the electrical permitivities of the intracavity medium and vacuum, respectively, and $c_0$ is the speed of light in vacuum. We shall mention that there exist other types of cavities beyond Fabry-Pérot, which achieve strong light-matter coupling with smaller number of molecules. These cavities are typically in the nanophotonic regime, but hereafter we will not be concerned with them, unless mentioned otherwise.

Next, one needs to keep in mind that the infrared light confined in these cavities mainly interacts with the molecular DOF related to nuclear displacement; therefore, under the cavity Born-Oppenheimer approximation (CBOA),\cite{Flick2017,Flick2017a} the EM mode is reasonably assumed as an additional nuclear coordinate corresponding to a vibrational mode with frequency $\omega_0$. We note that, since the cavity wavelength is several orders of magnitude larger than the molecular radius, the long-wavelength --or electric dipole-- approximation holds and it can be assumed that the EM mode interacts with a myriad of molecules simultaneously. Then, the description of the EM mode with phase space operators $\{\hat{p}_0,\hat{q}_0\}$ and its dipolar coupling with $N$ identical molecules is given by
\begin{equation}\label{eq:hamcav}
\hat{H}_\textrm{cav}=\frac{\hat{p}_0^2}{2}+\frac{1}{2}\qty(\omega_0\hat{q}_0+\sum_{i=1}^Ng(\hat{q}_0,\{\hat{R}_n\}_i))^2,
\end{equation}
with $g(q_0,\{R_n\}_i)=\sqrt{\omega_0}\chi\vb*{\epsilon}\cdot\vb*{\mu}(q_0,\{R_n\}_i)$ being the \emph{single-molecule} dipolar function, where $\vb*{\mu}(q_0,\{R_n\}_i)$ is the electric dipole moment of the $i$th molecule as a function of its nuclear configuration $\{R_n\}_i$, $\vb*{\epsilon}$ denotes the unitary EM field polarization vector, and $\chi=\sqrt{\varepsilon_0^{1/2}/\qty(2\pi \varepsilon^{3/2}c_0\mathcal{S})}$ is a coupling constant. \Cref{eq:hamcav} represents the cavity's contribution to the molecular ensemble, that is, none of these terms will be present when the molecular system is outside the cavity. Also, \cref{eq:hamcav} can only be applied to Fabry-Pérot cavities and does not accurately describe the case of nanophotonic cavities,\cite{Feist2021} an observation that has a few important consequences for vibropolaritonic chemistry, as discussed later. It shall be noted that \cref{eq:hamcav} relies on the Power-Zienau-Woolley transformation of the minimal coupling Lagrangian \cite{Vukics2021,Dupont1997,Landau1976}.

In terms of the normal-mode coordinates $\{q_{r\zeta}\}$, the \emph{single-molecule} dipolar function around a stationary point, indicated by $r$, of the PES can be written up to first order as
\begin{equation}\label{eq:dipfn}
g(q_0,\{q_{r\zeta}\})\approx g_{r}^{(0)}+\sum_{\zeta=1}^\Gamma g_{r\zeta}\rq{}q_{r\zeta}+\alpha_r\omega_0q_0,
\end{equation}
where $g_{r}^{(0)}=g(q_0=0,\{q_{r\zeta}=0\})$ and $g_{r\zeta}\rq{}=(\partial{g}/\partial{q_{r\zeta}})_{q_0=0,\{q_{r\zeta}=0\}}$ capture the roles of the permanent and transition dipole moments, respectively, while $\alpha_r$ denotes the static molecular polarizability, all evaluated at the stationary point $\{q_{r\zeta}=0\}$. We expect to have $N$ such dipolar functions, each representing the coupling betwen a specific molecule (with $\Gamma$ vibrational modes) and the cavity. As such, a vector-matrix notation is useful to write the ensuing expressions in a compact form. Let's define the $(N\Gamma+1)$-dimensional vector of coordinates as $\vb{q}=\qty(\tilde{q}_0,\vb{q}_{1}, \ldots,\vb{q}_{\Gamma})^T$, where $\tilde{q}_0=(1+\omega_0\chi\sum_{i=1}^N\alpha_i)q_0$ is a polarizability-scaled photon coordinate, and $\vb{q}_\zeta=\qty(q_{1,\zeta},\ldots,q_{N,\zeta})^T$ is the vector collecting the coordinates of the $\zeta$th normal mode across the $N$ molecules. Additionally, we have the $N\Gamma$-dimensional vector $\vb{g}\rq{}=\qty(\vb{g}'_{1},\ldots,\vb{g}'_{\Gamma}
)^T$, with $\vb{g}'_\zeta=(g'_{1,\zeta},\ldots,g'_{N,\zeta})^T$. Notice that the subscript $r$ indicating the stationary point has been replaced with the index $i$ that labels the molecule in the ensemble. As it will be seen, the former can be defined from the latter. To avoid further confusion in our notation, a Latin subscript labels the molecule, while a Greek letter labels the normal mode.

Recall that, to compute the TST rate of a single molecule system, we only need the PES up to second order in each $q_{r\zeta}$, that is, the only required information is in the neighborhood of the stationary points (see \cref{eq:rateconst} and \cref{eq:partfn}). With an ensemble of $N$ molecules, the PES should be described to $\order{\vb{q}^2}$. For instance, outside the cavity, the PES has the form $V_\textrm{mol}(\vb{q})=\vb{q}^T\qty(0\oplus\vb{A}_\textrm{mol})\vb{q}+c_\textrm{mol}+\order{\vb{q}^3}$, where $\vb{A}_\textrm{mol}$ is the diagonal matrix whose entries are the molecular frequencies, i.e. $\vb{A}_\textrm{mol}=\bigoplus_{\zeta=1}^\Gamma\vb{A}_{\zeta}$, with $\vb{A}_{\zeta}=\textrm{diag}(\omega_{1,\zeta}^2,\ldots,\omega_{N,\zeta}^2)$, and $c_\textrm{mol}=N\expval{V}_N$ is the potential energy at the stationary point. Here and hereafter, $\expval{x}_a=\sum_{i=1}^ax_i/a$ denotes averaging. Similarly, the contribution to the potential energy due to the cavity is given by \cref{eq:hamcav} and \cref{eq:dipfn} as $V_\textrm{cav}(\vb{q})=\vb{q}^T\vb{A}_\textrm{cav}\vb{q}+\vb{b}^T\vb{q}+c_{\textrm{cav}}+\order{\vb{q}^3},$ with coefficient arrays given by
\begin{align}
\vb{A}_\textrm{cav}=&\frac{1}{2}\begin{pmatrix}
\omega_0^2&\omega_0\vb{g}\rq{}^T\\
\omega_0\vb{g}\rq{}&\vb{g}\rq{}\vb{g}\rq{}^T
\end{pmatrix},\\
\vb{b}=&N\expval*{g^{(0)}}_N\begin{pmatrix}
\omega_0\\
\vb{g}'
\end{pmatrix},
\intertext{and}\label{eq:ccav}
c_\textrm{cav}=&\frac{N^2\expval*{g^{(0)}}_N^2}{2}.
\end{align}

\begin{figure}
\includegraphics[width=\linewidth]{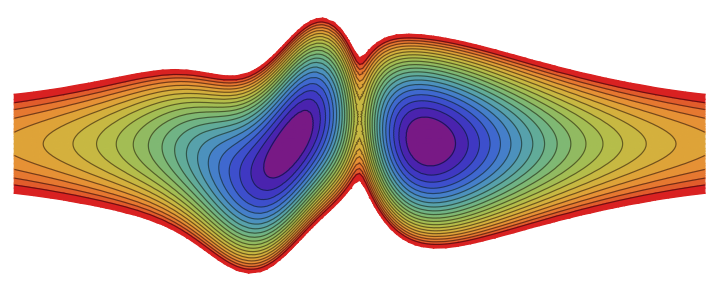}
\caption[Contour plot of a double-well PES where the reactant is coupled to the EM mode.]{Contour plot of double-well PES where the reactant is coupled to the EM mode. The horizontal axis corresponds to the reactive coordinate while the vertical axis to the electric field displacement. The frequencies of the eigenmodes can be read from the minor and major axes of the ellipses at the bottom of the wells. In the product case, the ellipse axes are aligned with the horizontal and vertical plot axes. In contrast, the ellipse in the reactant is tilted evidencing the formation of polaritons. This plot assumes that VSC is attained with a single molecule. \label{fig:coupes}}
\end{figure}

The Hamiltonian $\hat{H}=\hat{H}_\textrm{mol}+\hat{H}_\textrm{cav}$, customarily referred to as the Pauli-Fierz Hamiltonian\cite{Rokaj2018,Schaefer2020,Mandal2020}, describes the full system. The total potential energy is thus $V(\vb{q})=V_\textrm{mol}(\vb{q})+V_\textrm{cav}(\vb{q})=\vb{q}^T\vb{A}\vb{q}+\vb{b}^T\vb{q}+c+\order{\vb{q}^3}$, where $\vb{A}=\vb{A}_\textrm{cav}+(0\oplus\vb{A}_\textrm{mol})$, and $c=c_\textrm{mol}+c_\textrm{cav}$. We are now in a position to discuss the effects of $V_\textrm{cav}(\vb{q})$ on the total PES $V(\vb{q})$ (\cref{fig:coupes}). To begin with, the shift in the zeroth order term (in $\vb{q}$) granted by $c_\textrm{cav}$ implies that systems with non-vanishing permanent dipole moment will experience a generalized displacement of the energy reference. The presence of the first order term $\vb{b}$ suggests that, also for non-vanishing permanent dipole moments, the critical points (wells and saddles) may be relocated away from their original positions before VSC. Lastly, the contribution of $\vb{A}_\textrm{cav}$ implies that the bare cavity and molecular vibrational modes are coupled to each other and are no longer the eigenmodes. Instead, the coupled system is better described with new normal modes that mix the original modes and whose frequencies are shifted from the bare ones.

When calculating the stationary points for the coupled PES, it can be shown that none of the bare modes experience relocation of their respective critical points. On the other hand, the photon coordinate minimizes the coupled potential when $\tilde{q}_0=-N\expval*{g^{0}}_N/\omega_0$; consequently, the PES experiences no energy shift with respect to the uncoupled case. It shall be noted that this conclusion is contingent on the off-diagonal elements in the first row and column of $\vb{A}$ being proportional to $\omega_0$, as well as on the presence of the so-called self-interacting term $\vb{g}\rq{}\vb{g}\rq{}^T$. The resulting invariance is exact and independent of the coupling strength; therefore, it is more robust than what could be obtained from approximations that disregard these terms. Furthermore, the immobility of the stationary points is independent of whether the point includes an unstable component; therefore, the energy barrier remains the same with and without coupling.

Another relevant property of the matrix $\vb{A}$, also contingent on the self-interaction terms, is that
\begin{equation}\label{eq:freqprod}
\det(\vb{A})=\omega_0^2\det(\vb{A}_\textrm{mol})=\omega_0^2\prod_{i=1}^N\prod_{\zeta=1}^\Gamma\omega_{i\zeta}^2;
\end{equation} therefore, the product of eigenfrequencies is the same as the product of bare and cavity frequencies. This invariance implies that, if thermal equilibrium is assumed, the cavity promotes no energy exchange between the molecular modes, and the high-temperature partition function of the system [\cref{eq:partfn}] is the same with or without VSC. Other properties can be derived by considering that, given the structure of $\vb{A}_\textrm{cav}$, the matrix $\vb{A}_\textrm{mol}$ is the Schur complement of the $1\times1$ block $(\omega_0^2/2)$ in $\vb{A}$, i.e., $\vb{A}_\textrm{mol}=\vb{A}\backslash \qty(\omega_0^2/2)$.\cite{Liu2006}

Let's explore in more depth the specific situations that are typically regarded as relevant to explain kinetic modification.

We consider first the scenario where all $N$ molecules are in the RS, i.e. $r\to\textrm{eq}$ for $1\leq i\leq N$. Under this consideration, and assuming that the molecules can be treated as identical emitters (i.e., neglecting disorder), the potential can be recast as $V(\vb{q})=V_\textrm{B}(\vb{q}_\textrm{B})+V_\textrm{D}(\vb{q}_\textrm{D})+\order{\vb{q}^3}$, where $V_\textrm{B}(\vb{q}_\textrm{B})=\vb{q}_\textrm{B}^T\vb{A}_\textrm{B}\vb{q}_\textrm{B}+\vb{b}_\textrm{B}^T\vb{q}_\textrm{B}+c_\textrm{B}$ captures all the light-matter coupling effects, hence the subscript B for \emph{bright}, and $V_\textrm{D}(\vb{q}_\textrm{D})=\sum_{\zeta=1}^\Gamma(\tilde{\omega}_\zeta^2/2)\vb{q}_{\textrm{D}\zeta}^T\vb{q}_{\textrm{D}\zeta}+c_\textrm{D}$ is constituted by uncoupled eigenmodes with no information whatsoever from the photon coordinate, hence the subscript D for \emph{dark}. 

The dark modes are defined by the coordinates $\vb{q}_{\textrm{D}\zeta}=\vb{X}\vb{q}_\zeta$, where $\vb{X}$ is a $\qty(N-1)\times N$ matrix fulfilling $\vb{X}\vb{X}^\dag=\vb{I}_{N-1}$, and $\vb{X}\vb{g}_\zeta\rq{}=\vb{0}_{N-1}$, with $\vb{I}_{a}$ the $a$-dimensional identity matrix, and $\vb{0}_a$ the null vector with $a$ entries. Note that there are $\Gamma\qty(N-1)$ dark modes. Since these modes are degenerate, the constraints over $\vb{X}$ are not enough to define it unambiguously; therefore, the definition of the basis depends on the intended calculation. These modes retain the frequency of the bare ones and exclude any contribution from the photon mode; therefore, while not being formally identical to bare modes, they are assumed to behave as such. It is also possible to define bases such that the dark modes are \emph{almost} localized in a single bare mode. The zeroth order term for the dark potential is $c_\textrm{D}=\qty(N-1)V_\textrm{eq}$ and represents the potential energy of the dark modes at the stationary point.

The bright portion of the potential, on the other hand, is written in terms of the vector of coordinates $\vb{q}_\textrm{B}=\qty(\tilde{q}_0,q_{\textrm{B}1},\ldots,q_{\textrm{B}\Gamma})^T$, where $q_{\textrm{B}\zeta}=\vb{g}_\zeta\rq{}^T\vb{q}_\zeta/\mathpzc{g}'_{N\zeta}$ defines the $\zeta$th bright mode with normalization coefficient $\mathpzc{g}'_{N\zeta}=\sqrt{N\expval*{g_\zeta\rq{}^2}_N}$. The arrays of coefficients in the bright potential $V_\textrm{B}(\vb{q}_\textrm{B})$ are given by
\begin{align}\label{eq:brightarrays}
\vb{A}_\textrm{B}=&\frac{1}{2}\begin{pmatrix}
\omega_0^2&\omega_0\vb{g}_\textrm{B}\rq{}^T\\
\omega_0\vb{g}'_\textrm{B}&\vb{A}_\textrm{eq}+\vb{g}_\textrm{B}\rq{}\vb{g}_\textrm{B}\rq{}^T
\end{pmatrix},\\\label{eq:coupmatrs}
\vb{b}_\textrm{B}=&N\expval*{g^{(0)}}_N\begin{pmatrix}
\omega_0\\
\vb{g}'_\textrm{B}\\
\end{pmatrix},
\intertext{and}
c_\textrm{B}=&V_\textrm{eq}+c_\textrm{cav},
\end{align}
where $\vb{g}\rq{}_\textrm{B}=\qty(\mathpzc{g}\rq{}_{N1},\ldots,\mathpzc{g}\rq{}_{N\Gamma})^T$, and $\vb{A}_\textrm{eq}=\textrm{diag}(\omega_{\textrm{eq}1}^2,\ldots,\omega_{\textrm{eq}\Gamma}^2)$; again, $\vb{A}_\textrm{eq}=\vb{A}_\textrm{B}\backslash\qty(\omega_0^2/2)$. Note that the diagonal elements of the molecular portion in $\vb{A}_\textrm{B}$ are of the form $\tilde{\omega}_\zeta^2={\omega_{\textrm{eq}\zeta}^2+\mathpzc{g}_{N\zeta}\rq{}^2}$, thus corresponding to effective molecular frequencies.
It is important to highlight that the off-diagonal elements in $\vb{A}_\textrm{B}$ contain terms proportional to $\sqrt{N\expval*{g_\zeta\rq{}^2}_N}$. This implies that the photon mode interacts with the bright modes through collective couplings that are $\sqrt{N}$ times those due to the single-molecule transition dipole moments. Indeed, from experimentally observed Rabi splittings, the estimates for $N$ range between $10^{6}$ to $10^{12}$.\cite{Pino2015,Daskalakis2017} The form of these couplings reflects the collective nature of light-matter interaction as it should be clear that, since $g\rq{}_{1\zeta}\ll\omega_0$, the matrix $\vb{A}_\textrm{B}$ is effectively diagonal in the single-molecule limit, while $\vb{b}_\textrm{B}$ and $c_\textrm{cav}$ are negligible; consequently, there is no mixing between modes and no room for modification of the molecular properties.

To find the eigenmodes of $\vb{A}_\textrm{B}$, we assume in the following that $\omega_0\approx\omega_\Gamma$, that is, the cavity mode is near-resonant with the $\Gamma$th normal mode (coordinate $q_{\textrm{B}\Gamma}$), without loss of generality, while all other molecular normal modes (coordinates $q_{\textrm{B}1},\ldots,q_{\textrm{B}(\Gamma-1)}$) are far detuned from the cavity. This assumption implies that, effectively, only a single mode (here, coordinate $q_{\textrm{B}\Gamma}$) participates in the coupling while the others can be regarded as dark, which agrees well with experimental conditions. Then, within the latter "dark" off-resonant subspace, $\vb{A}_\textrm{B}$ is approximately diagonal (i.e. $\mathpzc{g}\rq{}_{N1}\approx\ldots\approx\mathpzc{g}\rq{}_{N(\Gamma-1)}\ll\mathpzc{g}\rq{}_{N\Gamma}$) and those $\Gamma-1$ "dark" modes are approximately eigenmodes of $\vb{A}_\textrm{B}$. All that remains is diagonalizing $\vb{A}_\textrm{B}$ in the subspace of the cavity and $\Gamma$th bright modes, a simple $2\times2$ matrix that affords an analytical solution to the remaining two eigenmodes known as the polariton modes. Overall, the bright potential is separated into $V_\textrm{B}(\vb{q}_\textrm{B})=V_+(q_+)+V_{-}(q_{-})+\sum_{\zeta=1}^{\Gamma-1}({\omega}_{\textrm{eq}\zeta}^2/2)q_{\textrm{B}\zeta}^2+V_\textrm{eq}$, with
\begin{equation}\label{eq:polpot}
V_\pm(q_\pm)=\frac{\omega_\pm^2}{2}q_\pm^2+b_\pm q_\pm+h_\pm^2c_\textrm{cav},
\end{equation}
where the subscripts $+$ and $-$ label, respectively, the upper and lower polariton modes. The polariton coordinates are $q_\pm=h_\mp q_{\textrm{B}\Gamma}\pm h_\pm\tilde{q}_0$, with corresponding frequencies $\omega_\pm=\sqrt{(\omega_0^2+\tilde{\omega}_\Gamma^2\pm\Omega^2)/2}$, where $h_\pm=\sqrt{\qty(1\pm\Delta^2/\Omega^2)/2}$ are the so-called Hopfield coefficients, $\Delta=\sqrt{\omega_0^2-\tilde{\omega}_\Gamma^2}$, and $\Omega^2=(\Delta^4+4\omega_0^2\mathpzc{g}_{N\Gamma}\rq{}^2)^{1/4}$. By analogy with the usual notation in CQED, the symbols $\omega_{\pm}$ and $\Omega$ denote the upper/lower polariton and Rabi frequencies, respectively, while $\Delta$ refers to the light-matter detuning; however, due to the presence of light-matter interaction terms beyond the rotating wave approximation, the relations among quantities in the current context are Pythagorean instead of linear.\cite{Flick2017a,Schaefer2018,Vendrell2018,Stokes2021} The coefficient for the linear term in \cref{eq:polpot} is given by $b_\pm=N\expval{g^{(0)}}_N\qty(h_\mp\mathpzc{g}'_{N\Gamma}\pm\omega_0h_\pm)$. It shall be noted that the experiments where VSC has resulted in modification of chemical kinetics were all conducted in liquid solutions; therefore, it is expected that the molecular orientations are isotropically distributed and, consequently, $\expval{g^{(0)}}_N=\sum_{i=1}^Ng^{(0)}_i/N$ should vanish. If this is the case, $b_\pm$ and $c_\textrm{cav}$  [\cref{eq:ccav}] become negligible and the only effect of the cavity on the shape of the PES is the mode renormalization with its ensuing frequency squeezing and swelling, i.e. the formation of new polariton modes with frequencies $\omega_\pm$ that differ from those of the cavity ($\omega_0$) and molecule ($\tilde{\omega}_\Gamma$) -- see \cref{eq:polpot}.

Chemical reactions are rare events\cite{Peters2017,Chandler1998}; therefore, the most accurate configuration describing the reacting system is that in which a single molecule is in the TS while all the others are in the RS. We explore this second scenario of interest next. Without loss of generality, we assume that the $N$th molecule is reactive, i.e., 
\[r\to\begin{cases}
\textrm{eq}&\text{for }1\leq i<N,\\
\ddag&\text{if }i=N,
\end{cases}\]
and, consistently with Sec. \ref{sec:TST}, we take the $\Gamma$th normal mode at the TS to be the unstable mode.

As, in general, $\omega_{\textrm{eq}\zeta}\neq\omega_{\ddag\zeta}$, having a molecule in the TS breaks the permutational symmetry that allowed the classification of the DOF presented above. However, it is enough to treat this single molecule in a separate fashion, and the $N-1$ remaining molecules will behave collectively as before. In fact, the treatment is the same until the definition of the dark and bright modes. Specifically, the dark subspaces per mode are of size $N-2$ instead of $N-1$, and the arrays of bright coefficients in \cref{eq:brightarrays} incorporate the single-molecule coupling to each of the normal modes. 
Under the assumption that all the stable modes in the reactive molecule are also far detuned from the cavity, and therefore reasonably removable from the coupling, the effective bright arrays to consider have the form
\begin{align}\label{eq:brightarraysTS}
\vb{A}_\textrm{B}^\ddag=&\frac{1}{2}\begin{pmatrix}
\omega_0^2&\omega_0\mathpzc{g}'_{N-1}&\omega_0g'_\ddag\\
\omega_0\mathpzc{g}'_{N-1}&\omega_\Gamma^2+\mathpzc{g}_{N-1}\rq{}^2&0\\
\omega_0g'_\ddag&0&\omega_\ddag^2+g_\ddag\rq{}^2
\end{pmatrix},\\\label{eq:coupmatts}
\vb{b}_\textrm{B}^\ddag=&N\expval*{g^{(0)}}_N\begin{pmatrix}
\omega_0\\
\mathpzc{g}'_{N-1}\\
g'_\ddag
\end{pmatrix},
\intertext{and}
c_\textrm{B}^\ddag=&V_\textrm{eq}+V_\ddag+c_\textrm{cav}.
\end{align}
Note that $g'_\ddag$ is a single-molecule coupling term whereas $\mathpzc{g}'_{N-1}=\sqrt{\qty(N-1)\expval*{g_\zeta\rq{}^2}_{N-1}}$ is a collective coupling term that is approximately $\sqrt{N-1}$ times larger than single-molecule couplings. In the large $N$ limit, $g'_\ddag\ll\mathpzc{g}'_{N-1}$. As such, the unstable (reactive) bare mode remains effectively dark, and all the polaritonic quantities are the same as before except with $N-1$ coupled modes instead of $N$. 
 
The above result is profound in that it implies that, when $N\gg1$, the stable modes at the equilibrium well are effectively indistinguishable from those at the saddle point; therefore, it is enough to describe the reaction with a single reacting molecule negligibly coupled to the confined EM mode. This discussion is consistent with the notion that chemical reactions are local phenomena. A parallelism can be drawn between molecular transformations and anharmonicites. As it has been shown, the local nature of anharmonic spectra makes them impervious to cavity effects \cite{Ribeiro2018} as they engage with the photon mode only as strongly as the single-molecule coupling allows.\cite{CamposGonzalezAngulo2022}

\subsection{Notions of vibropolaritonic chemical kinetics}
 
From the description in sec. \ref{sec:theopol}, we can summarize the main characteristics of VSC, and address some of the main misconceptions surrounding it.

First, VSC results in new normal modes: vibropolaritons and vibrational dark modes. In contrast with the typical picture inherited from excitonic strong coupling, where polaritons are regarded as eigenstates, VSC entails considering the EM field as an additional nuclear DOF. Second, all the molecules inside the mode volume contribute, upon proper alignment of their transition dipole moment, to the coupling. As a consequence, the modifications to the energy spectrum involve the whole molecular ensemble. In other words, the photon interacts with the bright mode, which is collective, to produce the polariton modes, but the contribution from each molecule to the bright mode is as large as its own transition dipole; therefore, the effect of the EM field on the molecule also depends on the single molecule dipole moment.  

With these considerations, let us warn about naive approaches to vibropolaritonic chemistry. It may be tempting to replace the reactive mode frequency with a polaritonic one; however, this strategy disregards that polariton modes do not align with the reactive coordinate, i.e., the reaction does not proceed along the polaritonic coordinates as these modes do not lead to the lowest energy barrier. In fact, a coherent vibration along a reactive bright mode takes the system to the extremely improbable configuration in which all coupled molecules react at the same time.

Another minimalist course of action is to consider that a single molecule couples to the cavity with an intensity afforded only collectively. This approximation is akin to solving the electronic structure of a polyelectronic neutral atom by comparing it to a hydrogenic system with a super electron whose charge is the negative of that of the whole nucleus. While intellectually interesting, and hypothetically insightful, this approach misrepresents the many body physics inherent to the problem. In technical terms, $N=1$ implies that there are no dark modes as the vector of coordinates is given by $\vb{q}_r=\qty(\tilde{q}_0,q_r)^T$, where it is also assumed that only one of the normal modes accurately describes the reactivity. Furthermore, $\mathpzc{g}'_N=g'_\textrm{eq}$; implicit in this identification is the fictitious enhancement of the single-molecule coupling constant so that VSC is artificially achieved in a Fabry-Pérot microcavity (alternatively, strong single-molecule light-matter coupling of $g'_{\textrm{eq}} \approx 0.01 \omega_{\textrm{eq}}$ could be achieved in nanophotonic resonators\cite{Chikkaraddy2016, Bitton2022}). Under such conditions, $g'_\ddag$ is no longer negligible and the reactive mode can no longer be assumed as dark; consequently, the properties of the polariton modes depend on the stationary point at which the reactive mode is found. To be specific, there are modes with frequencies $\omega_\pm^\textrm{eq}$ and $\omega_\pm^\ddag$ in the neighborhoods of the RS and TS, respectively. Additionally, $\expval{g^{(0)}}_1=g_{r}^{(0)}$, which means that permanent dipole moment effects are not quenched by isotropic averaging. Regardless of its shortcomings, this assumption is a decent starting point to inform chemical behavior, and as such, it is typically found as the first step in theoretical studies of vibropolaritonic chemistry.

\begin{figure}
\centering
\includegraphics[width=\linewidth]{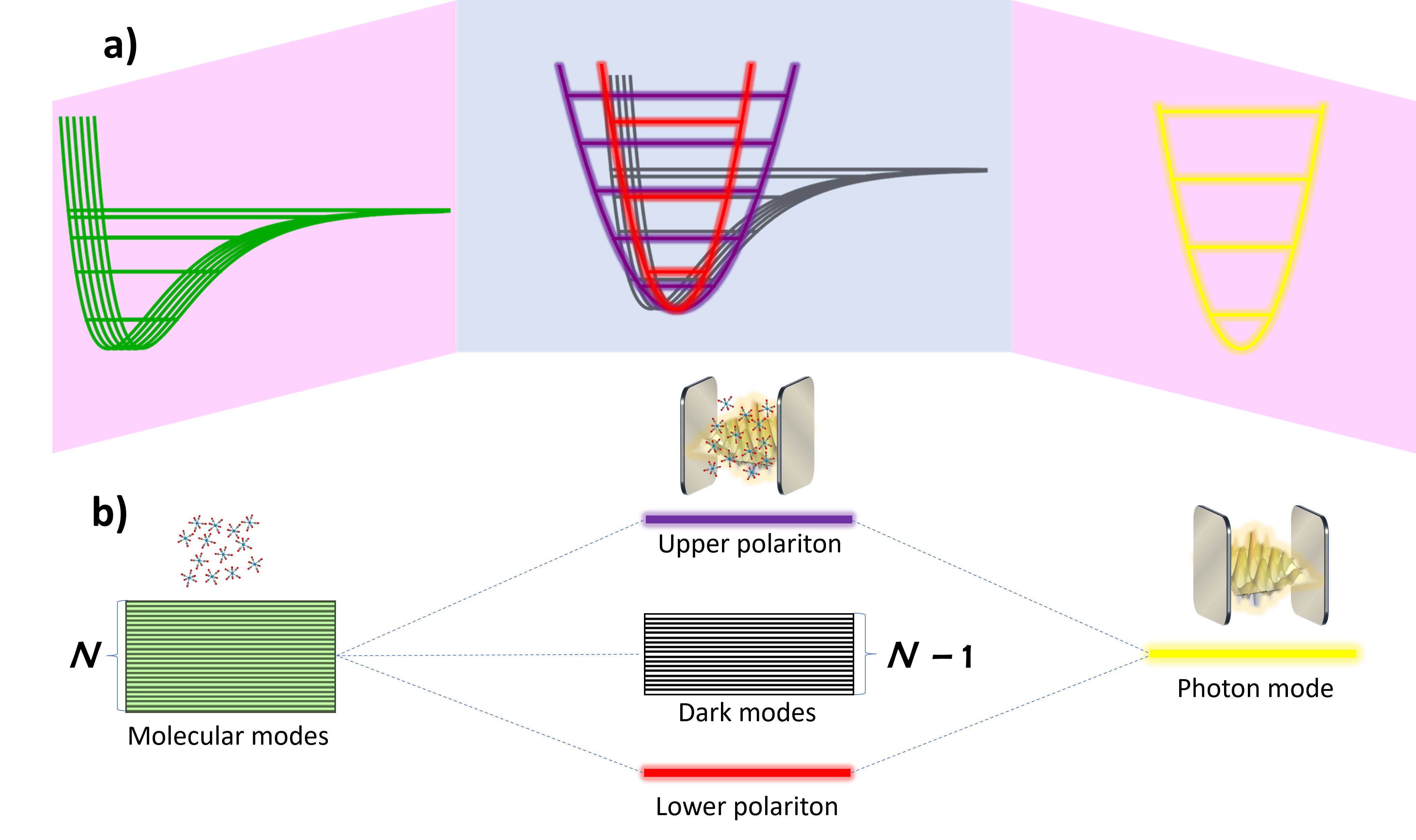}
\caption{Scheme describing $N$ molecular modes interacting strongly with a photon mode to give two polariton modes and $N-1$ dark modes. (a) Representation of the interaction at the PES level. (b) Correlation diagram corresponding to the singly-excited vibrational energy levels.
\label{fig:darksplit}}
\end{figure}

Following the formulation in sec. \ref{sec:theopol} the arguments that make vibropolaritonic chemistry counterintuitive can be summarized as follows:
\begin{itemize}
\item The presence of a large number of dark modes, as compared to only two polariton modes (\cref{fig:darksplit}), clearly indicates that thermodynamic averages at room temperatures are governed by the dark pure-molecular bundle, suggesting that most observables at equilibrium will be indistinguishable whether measured inside or outside of a resonator.
\item A chemical transformation can be regarded as the transition between configurations from $N$($\gg1$) to $N-1$ coupled molecules. Therefore, given that the individual light-matter couplings are very small, the impact of this change in the PES is negligible in the coupled ensemble.
\end{itemize}

\section{Theoretical approaches to VSC modified chemical kinetics\label{sec:approaches}}

In what follows, we give a nearly chronological recount of the theoretical efforts taken to understand vibropolaritonic chemistry.

The Feist group at the Universidad Autónoma de Madrid gave the first step into figuring out the implications of the presence of the cavity mode on chemical kinetics from a theoretical standpoint. Galego, Climent and collaborators worked through a hierarchy of formalisms, starting from a full-quantum first-principles description to validate the CBOA, and studied the changes to the rate derived from the modifications to the PES due to the additional photonic mode.\cite{Galego2019} In their derivation, the authors neglected the self interaction term; as a consequence, the squared molecular frequencies in \cref{eq:coupmatrs,eq:coupmatts} are not shifted by $g_r\rq{}^2$. Their findings can be summarized in the transmission coefficient
\begin{equation}\label{eq:feist}
\kappa_{\Delta E_a}=\exp(-\frac{u_\ddag-u_\textrm{eq}}{k_\textrm{B} T}),
\end{equation}
where $u_r=\qty(g_r \omega_0\omega_r/\omega_{+}^r\omega_{-}^r)^2$.\cite{CamposGonzalezAngulo2019}

According to this result, a chemical reaction can experience acceleration or suppression depending on the properties of the system at the stationary points. Specifically, \cref{eq:feist} represents an effective change in the activation energy (and thus the rate) of the reaction. Furthermore, the same group showed that, within this approach, the molecularity of the TS could be modified, as suggested by the thermodynamic parameters extracted from experiments.\cite{Climent2019} There are, however, some substantial setbacks to this framework. First, while neglecting the self-interaction is a customary practice in CQED for values of Rabi splitting in the range of those reported in experimental papers, the necessity of this term has been repeatedly shown in more recent discussions.\cite{Schaefer2020,Taylor2020,Stokes2021} As discussed in Sec. \ref{sec:theopol}, taking the self-interaction into account leaves the energies and nuclear configurations at the critical points (wells and saddle) of the PES unchanged regardless of the light-matter coupling strength; therefore, the change in activation energy should completely vanish.\cite{Li2021a} It is worth noting that several works\cite{Galego2019, Climent2019, CamposGonzalezAngulo2020} overlooked the dependence of the mode volume on the EM wavelength [\cref{eq:modevol}] and assumed that the terms $u_r$ are independent of the EM frequency, and thus concluded that resonance effects were impossible to observe. Although later studies have incorporated the frequency dependence, the maximization of the kinetic effect at resonance is still unapparent. Finally, the framework in Refs. \onlinecite{Galego2019, Climent2019} requires a non-vanishing permanent dipole moment at the stationary points for the reshaping of the PES they describe. Such a constraint makes this mechanism negligible in a collective ensemble of isotropically distributed dipoles.\cite{CamposGonzalezAngulo2020,Zhdanov2020}

Elaborating on the work in Ref. \onlinecite{Galego2019}, Kansanen and Heikkilä at the University of Jyväkylä extended Galego and collaborators\rq{} results by including detailed balance in the formulation, thus explicitly acknowledging the PS by studying a double-well PES.\cite{Kansanen2022} This work suggests that, when the transition dipole moment in the RS and PS has opposite signs, the coupling to the cavity induces a stability bifurcation on the distribution of states at equilibrium, i.e., the EM mode becomes a driving force on top of the energy difference between RS and PS that modifies the population bias at equilibrium. Unfortunately, this result depends on the reshaping of the cavity due to neglecting self-interaction, and it requires anisotropical alignment of emitters.

So far we have seen that the cavity leaves the activation barrier unchanged if we include the self-interaction term in the Hamiltonian. Since reaction rates are generally proportional to the exponential of the activation barrier [\cref{eq:rateconst} and \cref{eq:MLJrate}], the next logical step is looking at the prefactor of the exponential. For a system under light-matter coupling, as the one so far described, the transmission coefficient derived from mode renormalization can be written as $\kappa_Q=Q_\textrm{vib}^{-1}(\omega_{+}^\textrm{eq})Q_\textrm{vib}^{-1}(\omega_{-}^\textrm{eq})/Q_\textrm{vib}^{-1}(\omega_{+}^\ddag)Q_\textrm{vib}^{-1}(\omega_\textrm{eq})$.\cite{CamposGonzalezAngulo2020} Notice that, since $\omega_{-}^{\ddag2}<\omega_\ddag^{2}<0$, the lower polariton at the TS corresponds to the unstable eigenmode. At high temperatures, $\kappa_Q\approx\omega_{+}^\textrm{eq}\omega_{-}^\textrm{eq}/\omega_{+}^\ddag\omega_\textrm{eq}$, where $\omega_\textrm{eq}$ in the denominator cancels the prefactor of the bare rate. Furthermore, following \cref{eq:freqprod}, $\omega_{+}^\textrm{eq}\omega_{-}^\textrm{eq}=\tilde{\omega}_0^\textrm{eq}\omega_\textrm{eq}$, and $\kappa_Q\approx\tilde{\omega}_{0}^\textrm{eq}/\omega_{+}^\ddag$.

The Huo group at the University of Rochester noted that including the self-interaction makes the EM mode amenable to be treated as part of a harmonic bath. Following Pollak\rq{}s prescription for the Langevin equation,\cite{Pollak1986} the transmission coefficient can be rewritten as
\begin{equation}\label{eq:huo}
\kappa_\textrm{GH}=\frac{\omega_{-}^\ddag}{\omega_\ddag},
\end{equation}
where the subscript GH highlights the reminiscence between this expression and that of the Kramers-Grote-Hynes theory.\cite{Grote1980} Within Li and co-worker's formulation, the cavity mode can be interpreted as equivalent to a solvent coordinate and, thus, its effects on the kinetics is a consequence of dynamical caging,\cite{Li2021a} i.e., the EM mode stabilizes the nuclear configuration at the TS, preventing the system from reaching the PS. Additionally, changes in the free energy of activation calculated with this formalism follow similar trends with the coupling intensity as the experimentally observed. Moreover, in this approach, the transmission coefficient not only depends on the photonic frequency $\omega_0$ but also has a minimum at $\omega_0=\sqrt{\abs{\omega_\ddag}^2+{\eta'_\ddag}^4}-{\eta'_\ddag}^2$, where $\eta'_\ddag=g'_\ddag/2\sqrt{\omega_0}$ and $\abs{\omega_\ddag}$ is the frequency of the unstable mode; this relation affords an apparent rationalization of the experimentally reported mode-selective kinetic modifications.\cite{Thomas2019,Li2021} Since $\eta'_\ddag$ is generally small, there might be conditions such that  $\omega_0\approx\abs{\omega_\ddag}$ at the minimum of $\kappa_\textrm{GH}$, that is, mode selectivity occurs when the cavity is near-resonant with the unstable mode at its saddle point. Unfortunately, this resonant condition is not the one reported in experiments; instead, mode selectivities were observed when $\omega_0\approx\omega_\textrm{eq}$, with the latter term being the vibrational frequency near the equilibrium well. These are two very different conditions -- for instance, according to Ref. \onlinecite{Schafer2022a}, $\abs{\omega_\ddag}=\SI{74}{\per\centi\metre}$ in the deprotection reaction in Ref. \onlinecite{Thomas2016}; this frequency corresponds to only ca. $0.086\omega_\textrm{eq}$. More importantly, any modification to the rate according to this approach depends on the single-molecule light-matter coupling $g'_\ddag$ at the TS which, as we have seen earlier, is much smaller than the collective coupling $\mathpzc{g}'_{N-1}$ required to observe VSC.

Li and collaborators' work strongly suggests that, within the treatment so far outlined, the prefactor is independent of the reactive mode well frequency $\omega_\textrm{eq}$. One approach to correct the rate, so that the configuration at the equilibrium well is more relevant, is through the ZPE effects. The renormalization of modes introduced by the light-matter coupling produces a ZPE correction given by the transmission coefficient
\begin{equation}\label{eq:cao}
\kappa_\textrm{ZPE}=\exp(\frac{\hbar(\omega_{+}^\textrm{eq}+\omega_{-}^\textrm{eq}-\omega_{+}^\ddag-\omega_\textrm{eq})}{2k_\textrm{B}T}).
\end{equation}
According to this definition, the transmission coefficient can be written as 
\begin{equation}
\kappa_{Q}=\frac{\qty(1-\mathrm{e}^{-\hbar\omega_{+}^\textrm{eq}/k_\textrm{B}T})\qty(1-\mathrm{e}^{-\hbar\omega_{-}^\textrm{eq}/k_\textrm{B}T})}{\qty(1-\mathrm{e}^{-\hbar\omega_{+}^\ddag/k_\textrm{B}T})\qty(1-\mathrm{e}^{-\hbar\omega_\textrm{eq}/k_\textrm{B}T})}\kappa_\textrm{ZPE}.
\end{equation}
which allows to seamlessly separate its exploration at high and low temperatures. Such study was carried out by the Cao group at the Massachussetts Institute of Technology. Yang and Cao found that, at low temperatures whereby the ZPE contributions become relevant, it is possible to find minima in the transition coefficient at the experimentally observed resonant condition, $\omega_0=\omega_\textrm{eq}$, for various combinations of values of the transition dipole moment and unstable mode frequency.\cite{Yang2021} However, we need to keep in mind that ZPE effects are only observable at low temperatures.

In another attempt to bring the frequency at equilibrium into the picture, the Reichman group at Columbia University recently considered the Pollak-Graber-Hänggi (PGH) theory, a rate model which accounts for non-equilibrium effects in the low-friction regime.\cite{Pollak1989} The transmission coefficient derived from this approach is given by
\begin{equation}\label{eq:reichman}
\kappa_\textrm{PGH}=\frac{Y(\delta_\textrm{RS})Y(\delta_\textrm{PS})}{Y(\delta_\textrm{RS}+\delta_\textrm{PS})},
\end{equation}
where
\begin{equation}
Y(\delta_r)=\exp[\frac{1}{\pi}\int_{-\infty}^{\infty}\frac{\textrm{d}y}{1+y^2}\ln(1-\textrm{e}^{-\delta_r(1+y^2)/4})],
\end{equation}
and $\delta_r$ is the average energy loss that accounts for a) the friction due to the coupling to bath modes, and b) the force derived from the deviation of the PES with respect to the inverted parabola that describes the second order approximation of the TS, i.e., the presence of the equilibrium well at the RS. In the context of VSC, Lindoy and collaborators represented the cavity mode as an effective bath mode coupled to a single reactive mode and showed that this formulation successfully predicts resonant modifications to the rate.\cite{Lindoy2022} Subsequently, the Narang group at Harvard University and the Dou group at Westlake University showed in a collaborative study that, for imperfect cavities with multiple cavity modes, sharper resonances could be achieved.\cite{Philbin2022} However, all these works could only obtain a catalytic response. Moreover, Lindoy and co-workers discuss that it is unlikely for the low-friction conditions required by the theory to be met during the reported reactions. Importantly, these cavity effects also scale with the single-molecule light-matter coupling and are unlikely to explain those collective effects seen in VSC. The latter was investigated by Du and co-workers from the Yuen-Zhou group at UC San Diego, who extended the cavity PGH model to the collective regime.\cite{Du2022a} Consistently with sec. \ref{sec:theopol}, they described the effective bath modes as a pair of polariton modes formed by coupling of the cavity to $N-1$ other molecular modes. Unsurprisingly, the authors found that cavity effects occur mainly through the single-molecule coupling of the reacting molecule, while collective effects involving the $N-1$ remaining molecules are second-order in coupling since they occur through the cavity. Interestingly, Cao proposed that such cavity-induced frictions may be enhanced by spatial coherences within the molecular ensemble.\cite{Cao2022} However, apart from the aforesaid limitations which the author acknowledged, these coherence lengths are likely to be short in realistic systems.

Considering that the nuclear motion along the reaction coordinate is not harmonic, the Herrera group at the Universidad de Santiago de Chile studied the coupling between a photon mode and a single anharmonic oscillator.\cite{Hernandez2019,Triana2020} In the case of a Morse potential,\cite{Morse1929} Hernandez and coworkers found a shortening effect of the bond length. They interpret this observation as an increase in harmonicity of the excited states due to a significant photonic component. Later, Triana and collaborators concluded that, for more realistic molecular oscillators, the nuclear dynamics inside the cavity depends mostly on the shape of the dipole moment function; therefore, bond length enlargement and shortening are both possible. However, it has been shown that anharmonicities\cite{Ribeiro2018a,CamposGonzalezAngulo2021} are local as are chemical reactions; therefore, the interaction of the related degrees of freedom with the resonator is determined by the single-molecule transition dipole moment, which is negligible.

From the normal-mode analysis in Sec. \ref{sec:theopol}, it can be conjectured that TST cannot capture any effects from VSC. This point was formalized in a study by Li and coworkers at the University of Pennsylvania where they analyzed the potential of mean force (PMF) derived from the minimal coupling Hamiltonian.\cite{Li2020a} They found that the mean potential perceived by a single molecule due to the photon mode experiences a shift such that
\begin{equation}
V_\textrm{mol}(q)\to V_\textrm{mol}(q)+\frac{1}{2}\qty(\expval*{\hat{g}^2}_q-\expval*{\hat{g}}_q^2),
\end{equation}
where $\hat{g}=g(\hat{q}_0,\{\hat{R}_n\})$, and $\langle\hat{O}\rangle_q$ indicates the expectation value in the electronic ground-state of the operator $\hat{O}$ at the nuclear configuration described by the coordinate $q$. This shift is independent of $N$ and therefore incapable of capturing collective effects. Moreover, the orders of magnitude of the involved quantities make it negligible and thus unable to produce observable kinetic effects. The study also suggested contributions to the PMF of the $i$th molecule proportional to $\sum_{j\neq i}g_ig_j$, with $g_i=g(q_0,\{R_n\}_i)$. However, the authors discuss that this term can only be relevant for cavity lengths of the order of the intermolecular distances, or for anisotropically oriented samples. Furthermore, in the spirit of improving upon the classical approximations at the core of their work, the authors worked out a quantum path integral calculation. It remains unclear whether the PMF thus obtained reflects collective effects; nevertheless, given its analytical expression, it is unlikely that it captures resonant effects. 

The Rubio group at the Max Planck Institute assessed whether discarding the usual approximations incorporated in the minimal coupling Hamiltonian could provide more robust explanations of the modified kinetics. In Ref. \onlinecite{Hoffmann2020}, Hoffman and collaborators considered the interaction of a single molecule with multiple cavity modes. They found that the nuclear dynamics strongly changes as the number of considered photon modes increases. Adversely, this approach suffers from the same downsides as the one by the Herrera group when attempted in the collective regime. On the other hand, Schäfer and coworkers contrasted the implications of including and excluding the often-neglected self-interacting terms.\cite{Schaefer2020} They concluded that their absence leads to nonphysical situations; moreover, their inclusion might alleviate the incompatibilities between the local nature of a chemical reaction and the delocalized essence of light-matter coupling. Later on, Schäfer and collaborators also conducted a study simulating the trajectories of the reaction in Ref. \onlinecite{Thomas2016} using density functional theory (DFT) to describe the electronic structure of the involved molecules.\cite{Schafer2022a} The authors found that the cavity facilitates energy exchange between the intramolecular DOF, a process known as inter-mode energy redistribution (IMER); therefore, some of the kinetic energy allocated in the reactive coordinate can be redistributed among the other molecular modes, and it becomes harder for the system to reach the TS. The researchers also determined that energy exchange is more prominently activated when the EM mode is tuned to the frequencies of any of the normal modes in the bare molecule, in agreement with the experimental results in Ref. \onlinecite{Thomas2019}, as well as in the proximity of the barrier frequency; however, the calculated resonances are remarkably broader than those experimentally observed.

A similar study was carried out by the Vendrell group at the Universität Heidelberg, where they studied the isomerization of nitrous acid (HONO) by simulating classical trajectories on the PES calculated at the CCSD(T) level of theory.\cite{Sun2022} Sun and Vendrell also observed resonant intramolecular energy redistribution leading to rate modification. Remarkably, according to their calculations, the effect also leads to reaction acceleration. It is worth noting that, consistent with the PGH analysis, the presence of the cavity mode is only relevant in the under-damped regime. Alas, the authors performed simulations with several molecules participating in the coupling and found that the intermolecular energy exchange rate diminishes as the number of emitters grows, thus signaling not only that this is not a collective effect, but that with realistic values for the dipole moment function, it is unlikely that the phenomenon plays a role in reactions inside Fabry-Pérot resonators.

It is important to remark that the energy redistribution observed in the trajectory-based calculations cannot be captured by the multidimensional formulation of TST. The high-temperature limit of the transmission coefficient derived from considering all the normal modes is
the ratio of the product of mode frequencies with and without coupling. As previously pointed out, the structure of the coupling matrix makes it such that the determinants in both cases are identical, and the corresponding transmission coefficient is equal to 1.

It could be argued then that trajectory-based calculations have an edge since they highlight non-equilibrium effects that cannot be captured by TST.\cite{Sidler2022} Indeed, Wang and co-workers at Harvard University formulated a toy model for the dissociation of a triatomic molecule with bent geometry coupled to a cavity mode.\cite{Wang2022} Through analysis of classical trajectories, they also found out that the cavity induces intramolecular vibrational redistribution (IVR)\cite{Gruebele1998,Gruebele2004,Leitner2003,Leitner1997,Leitner2006}, leading to suppression of the dissociation rate. The authors also extended their study to the dissociation of a single Morse oscillator coupled to a cavity, and found that kinetic modification is possible if the dipole function includes terms beyond the linear approximation; however, their IMER based analysis did not show resonances with the fundamental mode.\cite{Mondal2022} Very recently, Wang and collaborators investigated collective ensembles of their toy molecule to assess whether the IMER mechanism they had observed in the single-molecule limit was present in the collective regime.\cite{Wang2022a} Their study indicates that this non-equilibrium effect may indeed be observed for a large number of emitters each weakly coupled to a common photonic mode. However, this effect is only valid for aligned dipoles, which brings into question its applicability in typical experimental settings.

In an interesting turn, Fischer and co-workers in the Saalfrank group at Universität Potsdam studied the isomerization of ammonia coupled to a cavity mode by solving the time-dependent Schrödinger equation. They not only confirmed the resonance effect with the barrier frequency as well as the ZPE cavity effects, but also discovered that, according to their quantum dynamics simulations using MCTDH, the EM mode itself can act as an energy gutter, thus suppressing the reaction even in the absence of additional modes available for IVR.\cite{Fischer2022} This is a truly resonant effect as the energy exchange depends on the tuning between the reactive and photon modes. 

In another attempt to address the contradiction between the non-locality of light matter coupling and the locality of chemical reactions, the Rubio group explored the possibility that collective effects could modify properties at the single-molecule scale. Making use of the CQED version of DFT the authors had developed, their work showed that an impurity in a polaritonic ensemble experiences electronic density modifications at the local level that translate into transition dipole changes increasingly scaling with the number of coupled molecules.\cite{Sidler2021} A similar phenomenon was suggested for the nonlinear response of silicon vacancy centers in diamond modeled with two-level systems.\cite{Botzung2020} According to Sidler and coworkers, the presence of the impurity results in an additional dark mode, or rather an effectively dark middle polariton, whose rearranged charge density may impact its chemical behavior. The local modifications persist even when the Rabi splitting is far larger than the detuning between the impurity and the group of identical emitters, which, the authors argue, mirrors the scenario of a reacting molecule among a collection of non-reacting ones. More recently, Schäfer provided strong support for this mechanism by computing the local chemical behavior of an impurity embedded in a polaritonic ensemble.\cite{Schafer2022} While certainly promising, this rationalization faces the limitations discussed on Ref. \onlinecite{Li2020a} where the intermolecular dipole-dipole interactions responsible for the local effects dwindle at equilibrium. Furthermore, in these studies, the transition dipole moment of the impurity is also artificially enhanced. In fact, Ref. \onlinecite{Schafer2022} discusses the role of the impurity with a Hopfield Hamiltonian reminiscent of \cref{eq:brightarraysTS}. The analogy thus leads to the same conclusion that, for large $N$, the impurity shall remain dark. Additionally, it is worth mentioning that the aforementioned study by Sun and Vendrell regarding the isomerization of HONO does not display effects consistent with this framework.\cite{Sun2022}

Following a similar rationale, the Huo group proposed a formulation in which the cavity is coupled to a solvent mode, therefore guaranteeing collective interaction, while the reactant couples dynamically to the solvent as prescribed by classical Kramers-Grote-Hynes theory.\cite{Mandal2022} This approach is also consistent with experimental results in similar settings.\cite{Lather2019,Lather2022} In that work, Mandal and coworkers calculated a reaction deceleration that increases with the amount of coupled emitters from the solvent and also presented the barrier resonance. The authors later accounted for cavity leakage, which leads to further slowdown of the reaction. However, their work includes the strong assumption that the reactive molecule interacts with the solvent mode in a way that mimics the light-matter coupling. In other words, the central hypothesis is that the solvation sphere for a single reactant includes the same number of molecules as those participating on the polaritonic ensemble, which is a questionable scenario. Furthermore, it has been reported that, in experimental settings with VSC to solvent, reaction acceleration is typically observed. Such a feature cannot be captured within this theoretical approach.

Thus far, we have presented works which describe chemical reactions as adiabatic transitions, i.e. works that apply the CBOA. Given that this description seems to prevent polaritonic phenomena from translating into kinetic effects, the Yuen-Zhou group looked into non-adiabatic processes.\cite{CamposGonzalezAngulo2019} Within this approach, the coupling between RS and PS is weaker than the light-matter interaction; therefore, the TS not only retains polaritonic features but is in fact defined by them. Campos-Gonzalez-Angulo (CGA) and coworkers formulated a VSC version of the MLJ theory. The coupling between high-frequency vibrations and the cavity mode affords modulation of some TSs present in this model. To summarize, for reactions that occur predominantly over the channel provided by the first vibrational excitation, the transmission coefficient derived with this approach has the form
\begin{equation}\label{eq:yuenzhou1}
\kappa_\textrm{NA}=\frac{h_{-}^2}{N}\textrm{e}^{-U_{+}^\ddag/k_BT}+\frac{h_{+}^2}{N}\textrm{e}^{-U_{-}^\ddag/k_BT}+\frac{N-1}{N},
\end{equation}
where $U_\pm^{\ddag}$ is the shift in TS state energy due to the difference in frequency between the bare and polariton modes, and $h_\pm$ is a Hopfield coefficient as defined in \cref{eq:polpot}. The relevance of this result relies in the fact that it makes explicit the interplay between entropic and enthalpic effects, the former given by the proportion of modes encoded in the prefactors, and the latter given by the TS energy shifts in the arguments of the exponentials. This approach offers an explanation for the modified kinetics as it is possible, in principle, that one of the first two terms in \cref{eq:yuenzhou1}, which include the polariton effects, surpasses the third one, which refers to the dark modes. This balance of terms has the interpretation that the boosted polaritonic mode opens a reactive channel with an energy barrier low enough for the reaction to prefer it over the vastly available dark channels. Furthermore, calculations show that this description is consistent with experiments in two accounts: the modification of the rate (a) increases with the number of coupled molecules, and (b) is maximal under resonant conditions.
For all its merits, the non-adiabatic formulation is not without caveats. For instance, significant changes to the TS will only produce $\kappa_\textrm{NA}>1$; therefore, deceleration effects are not captured. Moreover, there appear to be no substances with the reorganization energy required to achieve regimes of modified kinetics. More importantly, most of the systems for which experiments have shown kinetic modification have dynamics not accurately described within a non-adiabatic framework. Also, in the original work, it was the PS, instead of the RS, that coupled with the EM mode; however, the theory can be modified to include both without much qualitative change in the conclusions.

An additional possible flaw of the approach in Ref. \onlinecite{CamposGonzalezAngulo2019} was pointed out by Vurgaftman and collaborators at the U. S. Naval Research Laboratory in Ref. \onlinecite{Vurgaftman2020}. The authors of this analysis highlighted that transmission coefficients of the form $\kappa=\kappa_{+}+\kappa_{-}+\kappa_\textrm{D}$ cannot explain kinetic modifications for vibrational polaritons. To be specific the transmission coefficient needs to be rewritten to account for energy broadening effects as
\begin{equation}
\kappa=\int\textrm{d}\omega\qty[\rho_{+}(\omega)+\rho_{-}(\omega)+\rho_\textrm{D}(\omega)]\textrm{e}^{-U(\omega)/k_BT},
\end{equation}
where the distributions, $\rho_a(\omega)$, fulfill $\int\textrm{d}\omega\rho_\pm(\omega)=h_\mp^2/N$ and $\int\textrm{d}\omega\rho_\textrm{D}(\omega)=\qty(N-1)/N$, while the shifted TS energy, $U(\omega)$, is such that $U(\omega_D)=0$. Notice that in the work by Yuen-Zhou and coworkers $\rho_a(\omega)=\delta(\omega-\omega_a)$. In contrast, Vurgaftman and collaborators proved that,  given the energy scales of vibrational DOF, the broadening mechanisms are such that $\rho_D(\omega)\gg\rho_\pm(\omega)$ is always true under realistic experimental conditions, even at $\omega=\omega_\pm$, where $\rho_\pm(\omega)$ should be maximal. This relation between frequency distributions implies that the channels with lower activation energy are always available regardless of the presence of VSC, and the observable rate should remain unaffected by polaritonic effects.

With regards to the observation that rate modifications were maximal when the $k=0$ (normal incidence) cavity mode was resonant with molecular vibrational modes, this has not been well-explained thus far. Recall from Sec. \ref{sec:experim} that this finding is unexpected because red-detuning this cavity mode concomitantly shifts a $k>0$ (oblique incidence) cavity mode into resonance with the vibrational modes. In fact, Ribeiro at Emory University showed that the vibrational density of states, when inhomogeneously broadened, will overlap better with the cavity density of states if resonance occurs with a $k>0$ mode, thus increasing the polariton density of states.\cite{Ribeiro2022} However, an interesting insight was offered by Vurgaftman and co-workers in Ref. \onlinecite{Vurgaftman2022} by considering bulk polaritons when the molecular sample is placed outside the cavity. Relative to that, the polariton density of states was found to increase significantly when the molecules are moved into the cavity; moreover, this change is maximized when the $k=0$ mode is resonant with the vibrational mode. While certainly a step forward on the theoretical front, the authors also recognized that the computed polariton density of states remains a tiny fraction of the total vibrational density of states and cannot adequately account for the experimentally-observed resonance effect.

The same authors also provided new insights into why rate modifications were experimentally found to be maximal when the $k=0$ (normal incidence) cavity mode was resonant with molecular vibrational modes. (Recall from Sec. \ref{sec:experim} that this finding is unexpected because red-detuning this cavity mode concomitantly shifts a $k>0$ (oblique incidence) cavity mode into resonance with the vibrational modes. In fact, Ribeiro at Emory University showed that the vibrational density of states, when inhomogeneously broadened, will overlap better with the cavity density of states if resonance occurs with a $k>0$ mode, thus increasing the polariton density of states.\cite{Ribeiro2022}) None of the models mentioned thus far could explain the experimentally determined resonance condition, but what is unique about Ref. \onlinecite{Vurgaftman2022} is its consideration of bulk polaritons when the molecular sample is placed outside the cavity. Relative to that, the polariton density of states was found to increase significantly when the molecules are moved into the cavity; moreover, this change is maximized when the $k=0$ mode is resonant with the vibrational mode. While certainly a step forward on the theoretical front, the authors also recognized that the computed polariton density of states remains a tiny fraction of the total vibrational density of states and cannot adequately account for the experimentally-observed resonance effect.

In another work inspired by the non-adiabatic approach, Phuc and collaborators at the Institute for Molecular Science, in Okazaki, Japan, explored whether ultrastrong light-matter coupling could play a significant role as the formulation in Ref. \onlinecite{CamposGonzalezAngulo2019} was carried out under the rotating-wave approximation for which the vibrational ground-state of the coupled and bare systems are the same, i.e., $\ket{\nu_{+}=0,\nu_{-}=0}_r=\ket{\nu_0=0,\nu_r=0}$, where $r$ labels RS or PS, and $\nu_a$ indicates the vibrational excitation in mode $a$. Upon inclusion of the counterrotating and self-interaction terms (required in the ultrastrong coupling regime), the vibrational polaritonic ground-states are $\ket{\nu_{+}=0,\nu_{-}=0}_r=\sum_{\nu_0,\nu_r}\rq{}\xi_{\nu_0,\nu_r}^r\ket{\nu_0,\nu_r}$, where the sum, $\sum_{\nu_0,\nu_r}\rq{}$, is constrained to $\nu_0+\nu_r\in\text{ evens}$.\cite{Phuc2020} The authors found an impact on electron-transfer reactions dominated by the channel between reactant and product ground-states. The corresponding transmission coefficient has the form
\begin{equation}\label{eq:phuc}
\kappa_\textrm{US}=\frac{\abs{\sum_{\nu_0,\nu_\textrm{RS},\nu_\textrm{PS}}\rq{}\xi_{\nu_0,\nu_\textrm{RS}}^{\textrm{RS}}\xi_{\nu_0,\nu_\textrm{RS}}^\textrm{PS}F^{(0,\nu_\textrm{PS})}_{(0,\nu_\textrm{RS})}}^2}{\abs{F^{(0,0)}_{(0,0)}}^2}.
\end{equation}
As the coupling grows, the contribution from the Franck-Condon factor, $F^{(0,0)}_{(0,0)}$, decreases. Moreover, as the authors discuss, the coefficients $\xi_{00}^r$ and $\xi_{02}^r$ have opposite signs; therefore, the value of $\kappa_\textrm{US}$ decreases with the strength of the single molecule coupling. Nevertheless, the terms $\xi_{02}^r$ are proportional to $1/N$, which means that effects from this mechanism are negligible in the collective regime.

Facing the possibility that equilibrium properties of the polaritonic modes may not be responsible for the modified kinetics, the Yuen-Zhou group also looked into non-equilibrium effects involving the dark modes. In Ref. \onlinecite{Du2021}, Du and coworkers exploited the platform provided by the MLJ model, which allows for back reactions and nonreactive relaxation channels. Working in the regime where reactive processes and nonreactive relaxation occur on similar time scales, the authors found an alternative reaction mechanism that takes into account the substantially different dissipative pathways available to molecules inside a cavity versus outside. Three reaction schemes were studied:
\begin{subequations}
\begin{equation}
\schemestart
$A$
\arrow(R--P0){<=>[$k_1$][$k_{-1}$]}{$B^*$}
\arrow(@P0--P){->[$\gamma$]}$B$
\schemestop
\end{equation}
\begin{equation}
\schemestart
$A^*$
\arrow(R--P1){->[$k$]}$B$
\arrow(@R--R1){<=>[*0$\gamma_1$][*0$\gamma_{-1}$]}[-90]$A$
\schemestop
\end{equation}
and
\begin{equation}
\schemestart
$A$
\arrow(R--P1){->[$k_1$]}$B^*$
\arrow(@P1--P2){->[$k_2$]}$C^*$
\arrow(@P1--P10){->[*0$\gamma_1$]}[-90]$B$
\arrow(@P10--P20){->[$k_0$]}$C$
\arrow(@P2--P20){->[*0$\gamma_2$]}[-90]
\schemestop,
\end{equation}
\end{subequations}
where the asterisk indicates species in a vibrationally excited state amenable to polaritonic coupling, the rate constants $k$ characterize transitions between distinct nuclear equilibrium configurations, and the constants $\gamma$ parametrize the vibrational relaxations. For two molecules coupled to the cavity, it was found that, if the activation energies are insufficiently changed by polariton formation, cavity leakage significantly alters reaction kinetics. Specifically, cavity leakage provides a faster channel for vibrational decay and, by detailed balance, its reverse process. These new channels influence reactive events involving vibrational excited states. The first and second reaction schemes experience acceleration because cavity leakage effectively speeds up the transitions $B^* \rightarrow B$ and $A \rightarrow A^*$, respectively. In contrast, the third reaction scheme undergoes suppression of product formation, since cavity leakage promotes $B^*\rightarrow B$ over $B^* \rightarrow C^*$. However, as with other polaritonic effects, changes in reactivity due to cavity leakage are expected to vanish under collective VSC, when essentially all eigenmodes are dark. 

When inhomogeneous broadening is brought into consideration, ambiguities in the definition of the dark modes vanish and a unique basis is chosen such that a given molecule in the ensemble is mainly delocalized in just a few of the eigenmodes.\cite{Du2022} Thus, the value of each $k$ is reduced as the Franck-Condon factors involved in their calculation account for the distribution of the relevant mode among the dark and bright modes. On the other hand, given that the vibrational ground state is global and unique for all modes, the vibrational relaxation rates, $\gamma$, remain unaffected by the composition of the excited eigenstates. For the first reaction scheme, the transmission coefficient is
\begin{subequations}\label{eq:yuenzhou2}
\begin{equation}
\kappa_{ABB}=\qty(k_{-1}+\gamma)\sum_{i=1}^{N+1}\frac{\phi_i}{\phi_ik_{-1}+\gamma},
\end{equation}
where $\phi_i$ is the fraction of the reactive molecule in the $i$th eigenmode. Simulations indicate that there is an appreciable enhancement of the reaction rate for this process. The remaining reaction schemes have ideal transmission coefficients given by
\begin{align}
\kappa_{AAB}=&\qty(k+\gamma_1)\sum_{i=1}^{N+1}\frac{\phi_i}{\phi_ik_+\gamma_1}\\
\intertext{and}
\kappa_{ABC}=&\sum_{i=1}^{N+1}\phi_i^2;
\end{align}
\end{subequations}
however, simulations show that they do not describe the long term behavior of these processes. Although distant from experimental settings, the rationale afforded by this inherently collective approach brings into the picture the possibility of dark modes, rather than polaritons, being the driving force for modified kinetics. This concept is appealing because any rate changes due to dark modes will benefit from broadening, which has so far been detrimental to previous models. Nonetheless, this effect vanishes for $N>10$.

\Cref{tab:theories} summarizes all the referenced studies along with the experimental conditions and observations addressed.

\begin{table*}
\renewcommand{\arraystretch}{2}
\caption[Theories of vibrational-strong-coupling-modified ground-state reactivity.]{Theories of vibropolaritonic chemistry. Labels $\mathit{C1}$, $\mathit{O1}$ and $\mathit{O2}$ indicate the experimental features presented in Sec. \ref{sec:intro}; all theories addressed condition $\mathit{C2}$ (thermal reactions), and none addressed observation $\mathit{O3}$ (resonance with $k=0$).
\label{tab:theories}}
\begin{tabular}{lcccccl}
Mechanism&$k_\textrm{VSC}/k_\textrm{bare}$&Adiabatic&\makecell{Applicable to\\\textit{collective} VSC\\($\mathit{C1}$)}&\makecell{Effect\\on rate\\($\mathit{O1}$)}&\makecell{Frequency condition\\ for maximum effect\\($\mathit{O2}$)}&\makecell[c]{Notes}\\
\hline
\makecell[l]{Permanent dipole-induced\\changes in activation energy\cite{Galego2019,Climent2019,CamposGonzalezAngulo2020,Zhdanov2020}}&\cref{eq:feist}&\pmb{\checkmark}&\makecell[c]{With anistropic\\ alignment}&$\updownarrow$&\ding{55}&\makecell[l]{Vanishes when self-interaction\\terms are included\cite{Li2021a}}\\
Dynamical photon caging\cite{Li2021a,Li2021,Mandal2022}&\cref{eq:huo}&\pmb{\checkmark}&With solvent&$\downarrow$&$\omega_0\approx\abs{\omega_\ddag}$&\\
Zero-point energy\cite{Yang2021}&\cref{eq:cao}&\pmb{\checkmark}&\ding{55}&$\updownarrow$&$\pmb{\checkmark}$&\makecell[l]{Restricted to low temperatures\\Resonance condition not inherent}\\
Cavity-induced friction\cite{Lindoy2022,Philbin2022,Du2022a}&\cref{eq:reichman}&\pmb{\checkmark}&\ding{55}&$\uparrow$&$\omega_0=\omega_\textrm{RS}$&Requires low friction\\
\makecell[l]{Charge transfer\\(polaritons)\cite{CamposGonzalezAngulo2019,Vurgaftman2020}}&\cref{eq:yuenzhou1}&\ding{55}&\pmb{\checkmark}&$\uparrow$&$\omega_0\approx\omega_\textrm{PS}=\omega_\textrm{RS}$&Vanishes with broadening\\
\makecell[l]{Counter-rotating effects\\on groundstate\cite{Phuc2020}}&\cref{eq:phuc}&\ding{55}&\ding{55}&$\downarrow$&$\omega_0=\omega_\textrm{RS}$&\makecell[l]{Both RS and PS must couple\\to the cavity}\\
\makecell[l]{Charge transfer\\(dark modes)\cite{Du2021,Du2022}}&\cref{eq:yuenzhou2}&\ding{55}&\pmb{\checkmark}&$\updownarrow$&\pmb{\checkmark}\\
{IMER\cite{Wang2022,Mondal2022,Schafer2022a,Sun2022,Fischer2022,Wang2022a}}&N/A&\pmb{\checkmark}&\makecell[c]{With anisotropic\\alignment (?)}&$\updownarrow$&$\omega_0=\omega_\textrm{RS}$&\makecell[l]{Nonequilibrium\\ initial conditions\\Collective effects seen\\ in ref. \onlinecite{Wang2022a} but not in ref. \onlinecite{Sun2022}}
\end{tabular}
\end{table*}

\section{Discussion and outlook \label{sec:conc}}

From the compilation of works presented above, several lessons merit highlight. Formulations based on classical quasi-equilibrium have disagreed with experimental observations; in fact, TST asserts that no change in chemical kinetics should be observed due to the confined EM mode.\cite{CamposGonzalezAngulo2020,Zhdanov2020,Li2020a} In contrast, dynamics simulations seem to offer a plausible explanation of VSC modified kinetics in terms of IMER,\cite{Schafer2022a,Fischer2022,Sun2022,Wang2022,Wang2022a,Mondal2022} which can be regarded as a form of IVR that includes the photon mode. There are, however, valid reasons to remain skeptical about this mechanism. For instance, with regards to VSC, there seems to be conflicting results about the prevalence of this effect in the collective regime: while Wang and co-workers' toy molecules display IMER as the ensemble grows,\cite{Wang2022a} Sun and Vendrell's HONO molecules do not,\cite{Sun2022} even with aligned dipoles (which was a requirement by Wang).

Non-equilibrium considerations appear to be crucial as they offer a way to bypass the perceived obstacle that the dark modes represent.\cite{Du2021,Du2022} Nonetheless, it is not clear why any non-equilibrium effects would be relevant for reactions as slow as silane deprotection. In fact, the timescales for cavity leakage, IVR, and most other dynamical effects are short enough to be averaged out by the time taken for the reaction to occur.\cite{Berne1998}

The main lesson from the non-adiabatic picture seems to be the need for a state-to-state scheme;\cite{CamposGonzalezAngulo2019,Du2021} therefore, the RRKM rate theory emerges as an obvious alternative.\cite{Rice1927,Marcus1952} However, as soon as the formulation is brought from the microcanonical into the canonical ensemble, it transforms into TST, thereby leading to the same conclusions.

Importantly, none of the theoretical studies reported thus far has addressed the observation that VSC with the $k=0$ mode is necessary for rate modifications to occur. Whether this is an important detail to include in the models remains to be seen, although some preliminary evidence \cite{Tichauer2021,Ribeiro2022} indicates that this might indeed be the case.

Arriving to a functional explanation of VSC modified kinetics has become a transcendental pursuit as it not only offers to revolutionize practices in synthetic chemistry, creating a new paradigm for environmentally-friendly catalysis, but also provides a platform to refine well-established formulations of both light-matter interaction and chemical dynamics. In line with Ref. \onlinecite{Wang2021}, we encourage the experimental community to also report observations in which VSC fails to produce any changes to reactivity. It is our hope that a classification of the susceptibility of various chemical transformations to vibropolaritonic modification will reveal the patterns that theoreticians require to develop more successful approaches.

\begin{acknowledgments}
J.A.C.-G.-A. was supported through the AFOSR award FA9550-18-1-0289. M.D. and Y.R.P. were supported by the American Chemical Society Petroleum Research Fund for this research through the ACS PRF 60968-ND6 Grant. J.Y.Z. was supported by a Research Fellowship from the Alfred P. Sloan Foundation under Grant FG-2021-15653. J.A.C.-G.-A. thanks Alán Aspuru-Guzik for his support through the Canada Industrial Research Chairs Program.
\end{acknowledgments}

\section*{Data availability}
Data sharing is not applicable to this article as no new data were created or analyzed in this study.


%

\end{document}